\documentclass[a4paper]{jpconf}
\usepackage{graphicx}
\newcommand {\beq}{\begin{equation}}
\newcommand {\eeq}{\end{equation}}
\newcommand {\bea}{\begin{eqnarray}}
\newcommand {\eea}{\end{eqnarray}}
\newcommand {\nn}{\nonumber \\}
\newcommand {\m}{\mu}
\newcommand {\n}{\nu}
\newcommand {\pl}{\partial}

\newcommand {\al}{\alpha}
\newcommand {\be}{\beta}

\newcommand {\Ga}{\Gamma}
\newcommand {\x}{\xi}
\newcommand {\ka}{\kappa}
\newcommand {\la}{\lambda}
\newcommand {\La}{\Lambda}

\newcommand {\om}{\omega}

\newcommand {\ep}{\epsilon}

\newcommand {\na}{\nabla}
\newcommand {\del}  {\delta}
\newcommand {\Del}  {\Delta}
\newcommand {\mn}{{\mu\nu}}

\newcommand {\half}{ {\frac{1}{2}} }

\newcommand {\fourth} {\frac{1}{4} }
\newcommand {\Ecal}{{\cal E}}

\newcommand {\Lcal}{{\cal L}}

\newcommand {\Dcal}{{\cal D}}

\newcommand {\Wcal}{{\cal W}}
\newcommand {\gvec}{{\vec g}}

\newcommand {\vvec}{{\vec v}}
\newcommand {\xvec}{{\vec x}}

\newcommand {\ptil} {{\tilde p}}
\newcommand {\ktil} {{\tilde k}}
\newcommand {\ttil} {{\tilde t}}

\newcommand {\Lhat}{{\hat L}}

\newcommand {\delh} {{\hat \delta}}

\newcommand {\rdot}{\dot{r}}

\newcommand {\xdot}{{\dot{x}}}
\newcommand {\xddot}{\ddot{x}}
\newcommand {\bfZ} {{\bf Z}}
\newcommand {\K}{{\bf K}}
\newcommand {\I}{{\bf I}}

\newcommand {\intfx} {{\int d^4x}}

\newcommand {\intxy} {{\int d^4xdy}}

\newcommand {\intpL} {{\int_{\ptil\leq\Lambda} \frac{d^4p}{(2\pi)^4}}}
\newcommand {\intpE} {{\int \frac{d^4p_E}{(2\pi)^4}}}

\newcommand {\change} {\leftrightarrow}
\newcommand {\ra} {\rightarrow}

\newcommand {\pr}   {{\quad .}}
\newcommand {\com}  {{\quad ,}}
\newcommand {\q}    {\quad}

\newcommand {\qqqqq}   {\quad\quad\quad\quad\quad}
\newcommand {\qqqqqq}   {\quad\quad\quad\quad\quad\quad}

\newcommand {\nl}    {\newline}

\newcommand {\JHEP}  {{\it J. High Energy Phys.}}
\newcommand {\PTP}  {{\it Prog.Theor.Phys.}}
\newcommand {\Pla} {\frac{{\tilde p}}{\omega}}

\newcommand {\Tev} {\frac{{\tilde p}}{T}}

\newcommand {\Hora} {Ho\u{r}ava}      %090615
\begin{document}
\title{Geometric Approach to Quantum Statistical Mechanics and 
Application to Casimir Energy and Friction Properties}

\author{Shoichi Ichinose}

\address{
Laboratory of Physics, School of Food and Nutritional Sciences, 
University of Shizuoka\\
Yada 52-1, Shizuoka 422-8526, Japan
}

\ead{ichinose@u-shizuoka-ken.ac.jp}

\begin{abstract}
A geometric approach to 
general quantum statistical systems (including the harmonic oscillator) 
is presented. It is applied to Casimir energy and the dissipative system 
with friction. 
We regard the (N+1)-dimensional Euclidean {\it coordinate} system (X$^i$,$\tau$) 
as the quantum statistical system of N quantum (statistical) variables (X$^i$) and 
one {\it Euclidean time} variable ($\tau$). 
Introducing paths (lines or hypersurfaces) in this space (X$^i$,$\tau$), 
we adopt the path-integral method to quantize the mechanical system. 
This is a new view of (statistical) quantization of the {\it mechanical} system. 
The system Hamiltonian appears as the {\it area}. 
We show quantization is realized by  
the {\it minimal area principle} in the present geometric approach. 
When we take a {\it line} as the path, 
the path-integral expressions of the free energy are shown to be 
the ordinary ones (such as N harmonic oscillators) or their simple 
variation. When we take a {\it hyper-surface} as the path, 
the system Hamiltonian is given by the {\it area} of the {\it hyper-surface} 
which is defined as a {\it closed-string configuration} in the bulk space. 
In this case, the system becomes a O(N) non-linear model. 
We show the recently-proposed 5 dimensional 
Casimir energy (ArXiv:0801.3064,0812.1263) is valid. 
We apply this approach to the visco-elastic system, and present a new method 
using the path-integral for the calculation of the dissipative properties. 
\end{abstract}

\section{Introduction\label{intro}}
In the quest for the fundamental structure of the space, time, and matter, the most advanced 
theories are the string theory, D-brane theory and M-theory\cite{StringText}. 
They are beyond the quantum field theory in that the extended (in space-time) 
objects are treated as fundamental elements. Since the finding of AdS/CFT 
correspondence\cite{Malda9711,GKP9802,Witten9802}, various 
new ideas and techniques, developed for them, are imported 
into the {\it non-perturbative} analysis of the quantum field theories. In particular, 
the application to the material physics is marvelous: heavy ion collision 
physics and the  
viscosity in the quark-gluon plasma(\cite{Natsu0701,Son0704,Mateos0709} 
for review), superconductivity and superfluidity
\cite{Gubser0801,HHH0803,GuPu0805,HHH0810}, 
baryon mass spectrum in QCD\cite{SaSu0412,SaSu0507}. 
In this circumstance, a {\it new standpoint} about the space-time quantization 
appear. One is proposed by \Hora\cite{Hora0812,Hora0901}. He introduced Lifshitz's 
higher-derivative scalar theory and its renormalization group behavior into 
his idea about the new quantum gravity. Another one is revively given by E. Verlinde\cite{EVerlinde1001}. 
He emphasizes the entropic force (rather than energetic force) and the thermodynamical 
behavior near the horizon (Hawking radiation). 
With this recent trend of the geometrical view, the statistical(thermal) view
and the visco-elastic view 
, we present a {\it new} formalism where the quantum statistical system is treated 
purely in the geometrical way.     

In the analysis of the dissipative material-system, due to friction, one familiar way is to 
treat it as the {\it hydrodynamic} system of a continuum medium. The incompressible viscous flow 
(Newton' flow is assumed) is discribed by Navier-Stokes equation. 
%*** intro1%%%%%%%%%%%%%%%%
\bea
\frac{Dv^i}{Dt}\equiv (\frac{\pl}{\pl t}+v^j\pl_j)v^i
=-\frac{1}{\rho}\pl_iP+\frac{\eta}{\rho}\Del v^i+g^i\com\nn
i,j=1,2,3\com\q \xvec\equiv (x^1,x^2,x^3)=(x,y,z)\com\q
\pl_i\equiv\frac{\pl}{\pl x^i}\com\q \Del\equiv \pl_1^2+\pl_2^2+\pl_3^2\com\nn
v^i=v^i(t,\xvec)\com\q P=P(t,\xvec)\com
\label{intro1}
\eea 
%%%%%%%%%%%%%%%%%%%%%%%%%%%
where $\gvec\equiv (0,0,g)$ is the gravitational acceleration constant. 
$\rho$ and $\eta$ is the mass density (constant) and the viscosity (constant) of the fluid, 
respectively. $\vvec\equiv (v^1,v^2,v^3)$ is the velocity field and $P(t,\xvec)$ is the pressure field. 
This equation has been examined analytically and numericallly in various ways. 
The statistical quantities of the system has been obtained, for example, by 
introducing the random force (Langevin equation, the stochastic method, Fokker-Planck equation). 
In the present work, by introducing 
a {\it metric} in $(t,x^i)$-space, we treat this system geometrically. The statistical or global quantities 
(energy, entropy,...) are exprssed in the {\it path-integral} form. 

As another application, we take Casimir energy. It is the vcuum (zero temperature) 
energy for the free (non-interacting) part of the quantum system. It is a basic quantity 
of the quantum field theory (QFT) such as the field of radiation (the electromagnetic field). 
Generally it is expressed by the boundary parameters of the system. 
The quantity is so delicately defined that we need careful regularization of divergences. 
The present approach gives a new regularization which has some characteristic points compared 
with the ordinary one. Especially the direction of the {\it renormalization group} flow 
determines the attractiveness or repulsiveness. We also apply the present approach 
to 5 dimensional (dim) field theories (flat and curved ones), and well-define the higher-dimensional 
QFTs. 

The content is organized as follows. We start with 
the simple quantum statistical system of one harmonic oscillator in Sec.\ref{oneHO}. 
We see the geometric approach 
works well by regarding the {\it extra coordinate} as the {\it Euclidean time}. 
%**compell** us the bulk space 
%be more generalized than the ordinary metric-defined one. 
We generalize the harmonic oscillator potential (elastic system) to 
the general one in Sec.3. 
In Sec.4 the one variable system is generalized to the system of N variables.  
We analyze the quantum statistical system in the 
N+1 extra dimensional Euclidean {\it geometry}. 
The O(N) nonliner model 
naturally appears by taking  
the {\it closed-string} configuration. 
We stress that taking the area as Hamiltonian is 
one realization of the {\it minimal area principle}. 
In Sec.5, we apply the present approach to Casimir energy. 
The visco-elastic system is treated in Sec.6. 
We conclude in Sec.7. In Appendix A, some detail of Sec.5.2  
(Casimir energy in 5D curved space) is explained.

%%%%%%%%%%%%%%%%%%%%%%%%%%%%  Sec.2  %%%%%%%%%%%%%%%%%%%%%%%%%%%%%%%%%
%%%%  Quantum Statistical System of One Harmonic Oscillator  %%%%%%%%%
%%%%%%%%%%%%%%%%%%%%%%%%%%%%%%%%%%%%%%%%%%%%%%%%%%%%%%%%%%%%%%%%%%%%%%
\section{Quantum Statistical System of Harmonic Oscillator\label{oneHO}}
%***label***{oneHO}

\subsection{'Dirac' Type\label{1HOsup}}
\begin{figure}[h]
\includegraphics[width=14pc]{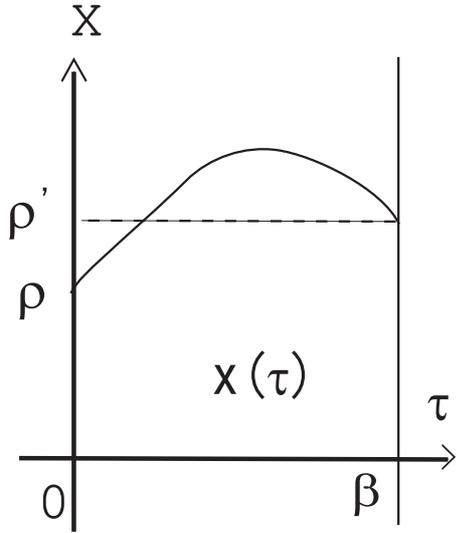}\hspace{2pc}%
\begin{minipage}[b]{14pc}\caption{\label{2DPath}
A path of line in 2D Euclidean space (X,$\tau$). The path starts
at x(0)=$\rho$ and ends at x($\be$)=$\rho'$. 
                                  }
\end{minipage}
\end{figure}

%***label***{1HOsup}\nl
Let us consider 2 dim Euclidean space $(X,\tau)$ described by the following metric. 
%*** oneHO1%%%%%%%%%%%%%%%%
\bea
ds^2=dX^2+\om^2X^2d\tau^2=G_{AB}dX^A dX^B\com\nn
(X^A)=(X^1,X^2)=(X,\tau)\com\q (G_{AB})=\mbox{diag}(1,\om^2X^2)\com\nn
R_{AB}=0 \com\q R=G^{AB}R_{AB}=0 \com
\label{oneHO1}
\eea 
%%%%%%%%%%%%%%%%%%%%%%%%%%%
where $A,B=1,2$. The parameter $\om$ is the 'spring' constant with the dimension of mass. 
We impose the periodicity (period: $\beta$) in the direction of the extra dimension $\tau$. 
%*** oneHO2%%%%%%%%%%%%%%%%
\bea
\tau\ra\tau+\be
\pr
\label{oneHO2}
\eea 
%%%%%%%%%%%%%%%%%%%%%%%%%%%
This is a way to introduce the {\it temperature} ($1/\be$) in the system. 
Here we take a path $\{ x(\tau),\ 0\leq \tau\leq \be\}$ in the 2D bulk space (X,$\tau$) and 
the {\it induced} metric on the {\it line} is given by
%*** oneHO3%%%%%%%%%%%%%%%%
\bea
X=x(\tau)\com\q 
dX=\xdot d\tau\com\q \xdot\equiv\frac{dx}{d\tau}\com\q
0\leq\tau\leq\be\com\nn
ds^2=(\xdot^2+\om^2x^2)d\tau^2
%\left[*********=(1+\frac{\om^2x^2}{\xdot^2})dx^2*******\right]
\pr
\label{oneHO3}
\eea 
%%%%%%%%%%%%%%%%%%%%%%%%%%%
See Fig.\ref{2DPath}. 
                             %%%   <Fig.2   %%%
                              %%%   Fig.2>  %%%

Then the {\it length} L of the path $x(\tau)$ is given by 
%*** oneHO4%%%%%%%%%%%%%%%%
\bea
L=\int ds=\int_0^\beta\sqrt{\xdot^2+\om^2x^2}d\tau
\pr
\label{oneHO4}
\eea 
%%%%%%%%%%%%%%%%%%%%%%%%%%%
We take the half of the length ($\half L$) as the system Hamiltonian 
({\it minimal length principle}). 
Then the free energy $F$ of the system is given by 
%*** oneHO5%%%%%%%%%%%%%%%%
\bea
\e^{-\be F}=\int_{-\infty}^{\infty}d\rho
\int_{\begin{array}{c}x(0)=\rho\\x(\be)=\rho\end{array}}
\prod_\tau\Dcal x(\tau)\exp \left[-\half\int_0^\beta\sqrt{\xdot^2+\om^2x^2}d\tau\right]
\com
\label{oneHO5}
\eea 
%%%%%%%%%%%%%%%%%%%%%%%%%%%
where the path-integral is done for all possible paths with the indicated boundary condition (b.c.). 
This quantum statistical system can be regarded as the square-root type ('Dirac' type) 
of the ordinary harmonic oscillator.
\footnote{
The situation reminds us of the relation between Nambu-Goto action 
and Polyakov action in the string theory\cite{Pol81}. 
The introduction of an auxiliary variable helps to 'normalize' 
the square-root action (\ref{oneHO5}). In this case, the geometric role of 
the auxiliary variable remains obscure.
} 
%*********This is the free energy of the {\it super} harmonic oscillator.  REFERENCES !! **************

\subsection{Standard Type\label{1HOnrm}}
\begin{figure}[h]
\includegraphics[width=14pc]{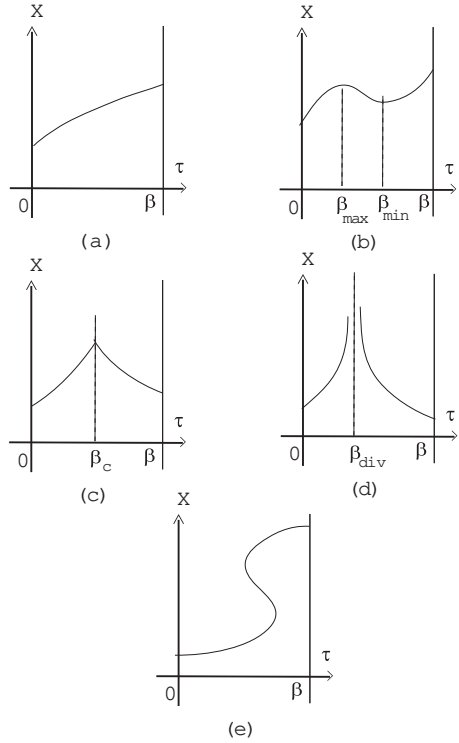}\hspace{2pc}%
\begin{minipage}[b]{14pc}\caption{\label{2DPaths}
Singular and regular lines in 2D Euclidean space (X,$\tau$).\ 
(a) regular line, simply increasing;\ 
(b) regular line, maximum at $\beta_{max}$ and minimum at $\beta_{min}$; 
(c) singular line, different derivatives for $\beta\ra\beta_c\pm 0$; 
(d) singular line, divergent at $\beta_{div}$; 
(e) singular line, multi-valued. 
%***2DPaths.eps\
                                  }
\end{minipage}
\end{figure}

%***label***{1HOnrm}\nl
Now we consider another type of 2 dim Euclidean space $(X,\tau)$ described by the following
{\it line} element. 
%*** oneHO6%%%%%%%%%%%%%%%%
\bea
ds^2=\frac{1}{d\tau^{2}}(dX^2)^2+\om^4X^4d\tau^2+2\om^2X^2dX^2 
=\frac{1}{d\tau^2}(dX^2+\om^2X^2d\tau^2)^2
\com
\label{oneHO6}
\eea 
%%%%%%%%%%%%%%%%%%%%%%%%%%%]
where we put the following condition on the infinitesimal quantities, $d\tau^2$ and $dX^2$,  
in order to keep all terms of (\ref{oneHO6}) in the same order. 
%*** oneHO7%%%%%%%%%%%%%%%%
\bea
\mbox{[Line Element Regularity Condition]}\ :\q\q\q\q\q\q\q\nn  
d\tau^2\sim O(\ep^2)\com\q dX^2\sim O(\ep^2)\com\q
\frac{1}{d\tau^2}dX^2 \sim O(1) 
\com
\label{oneHO7}
\eea 
%%%%%%%%%%%%%%%%%%%%%%%%%%%
where $\ep$ is an arbitrary infinitesimal parameter with the dimension of length. 
\footnote{
The condition (\ref{oneHO7}) 
restricts the trajectory configuration (\ref{oneHO9}) only to 
smooth-lines 
in the 2D bulk space, and excludes singular-lines which have some {\it singular} 
points (the derivative along $\tau$ can not be defined) between $0\leq\tau\leq\be$. 
See Fig.\ref{2DPaths} for singular and regular lines. 
%See the final section for the discussions about 
%an interpretation of the condition (\ref{oneHO7}). 
}
                             %%%   <Fig.3   %%%
                              %%%   Fig.3>  %%%
Note that we do {\it not} have 2D metric in this case. 
(We cannot define the bulk metric $G_{AB}(X)$.) 
We impose the periodicity (period: $\be$).
%*** oneHO8%%%%%%%%%%%%%%%%
\bea
\tau\ra\tau+\be
\pr
\label{oneHO8}
\eea 
%%%%%%%%%%%%%%%%%%%%%%%%%%%
Here we take a path $\{x(\tau),\ 0\leq \tau\leq \be\}$, and 
the {\it induced} metric on the line is given by
%*** oneHO9%%%%%%%%%%%%%%%%
\bea
X=x(\tau)\com\q 
dX=\xdot d\tau\com\q \xdot\equiv\frac{dx}{d\tau}\com\q
0\leq\tau\leq\be\com\nn
ds^2=(\xdot^2+\om^2x^2)^2d\tau^2
%\left[****=(\xdot+\frac{\om^2x^2}{\xdot})^2 dx^2****\right]
\pr
\label{oneHO9}
\eea 
%%%%%%%%%%%%%%%%%%%%%%%%%%%
In the bulk we do {\it not} have the metric, but 
on the path, we {\it do} have this {\it induced} metric. 
Then the {\it length} L of the path $x(\tau)$ is given by 
%*** oneHO10%%%%%%%%%%%%%%%%
\bea
L[x(\tau)]=\int ds=\int_0^\beta(\xdot^2+\om^2x^2)d\tau
\pr
\label{oneHO10}
\eea 
%%%%%%%%%%%%%%%%%%%%%%%%%%%
Hence, taking $\half L$ as the Hamiltonian ({\it minimal length principle}), 
the free energy $F$ of the system is given by 
%*** oneHO11%%%%%%%%%%%%%%%%
\bea
\e^{-\be F}=\int_{-\infty}^{\infty}d\rho
\int_{\begin{array}{c}x(0)=\rho\\x(\be)=\rho\end{array}}
\prod_\tau\Dcal x(\tau)\exp \left[-\half\int_0^\beta(\xdot^2+\om^2x^2) d\tau\right]
\com
\label{oneHO11}
\eea 
%%%%%%%%%%%%%%%%%%%%%%%%%%%
where the path-integral is done for all possible paths with the indicated b.c.. 
This is exactly the free energy of the harmonic oscillator. 
See Feynman's textbook\cite{Fey72}. 
\footnote{
$
F=\frac{\om}{2}+\frac{1}{\be}\ln(1-\e^{-\be\om}), 
E=<\frac{L}{2}>=\frac{\om}{2}\coth(\frac{\om\be}{2})=\frac{\om}{2}+\frac{\om}{\e^{\om\be}-1}, 
S=\frac{1}{T}(E-F)=k\{\frac{\be\om}{2}\coth\frac{\be\om}{2}-\frac{\be\om}{2}-
\ln(1-\e^{-\be\om}) \}
$
}

Note that the condition (\ref{oneHO7}) is necessary for the {\it elastic} 
view to the path. 
%%%%%%%%%%%%%%%%%%%%%%%%%%%%  Sec.3  %%%%%%%%%%%%%%%%%%%%%%%%%%%%%%%%%
%%%%                                                            %%%%%%
%%%%             General Quantum Statistical System             %%%%%%
%%%%                                                            %%%%%%
%%%%%%%%%%%%%%%%%%%%%%%%%%%%%%%%%%%%%%%%%%%%%%%%%%%%%%%%%%%%%%%%%%%%%%
\section{General Quantum Statistical System\label{GQSS}}
%***label***{GQSS}
We generalize the harmonic oscillator potential, $\half \om^2X^2$, to the 
general one $V(X)$.  
As for $V(X)$, we have the following form in mind. 
%*** GQSS0%%%%%%%%%%%%%%%%
\bea
\frac{\om^2}{2}X^2+\frac{\la_3}{3!}X^3+\frac{\la_4}{4!}X^4+\cdots
\com
\label{GQSS0}
\eea 
%%%%%%%%%%%%%%%%%%%%%%%%%%%
where $\la_3, \la_4, \cdots$ are the coupling constants for additional terms. 
%%%%%%%%%%%%%%%%%%%%%%%%%%%
\subsection{'Dirac' Type\label{GQSSsp}}  %***GQSSsp**
We start with the following metric in 2 dim Euclidean space $(X,\tau)$. 
%*** GQSS1%%%%%%%%%%%%%%%%
\bea
ds^2=dX^2+2V(X)d\tau^2=G_{AB}dX^A dX^B\com\nn
(X^A)=(X^1,X^2)=(X,\tau)\com\q (G_{AB})=\mbox{diag}(1,2V(X))\com\nn
(R_{AB})=\left(
\begin{array}{cc}
\frac{V''}{2V}-\fourth\left(\frac{V'}{V}\right)^2,  &  0\\
0\com    &  V''-\half\frac{(V')^2}{V}  
\end{array}
         \right) \com\q 
R=G^{AB}R_{AB}=\frac{V''}{V}-\half\left(\frac{V'}{V}\right)^2 \com\nn
V'\equiv \frac{dV(X)}{dX}\com\q V''\equiv \frac{d^2V(X)}{dX^2}\com
\label{GQSS1}
\eea 
%%%%%%%%%%%%%%%%%%%%%%%%%%%
where $A,B=1,2$. Note that V(X) does {\it not} depend on $\tau$. 
%\footnote{
%We furthermore note the new standpoint about the quantization of gravity (metric), in the present approach. 
%The most familiar way is to regard the 'metric' $V(X)$ as a {\it field} variable at the point ($X,\tau$) 
%and quantize it field-theoretically. We do not take such way of quantization. We accept the potential 
%form of $V(X)$ as a given one (background treatment) and do not treat $V(X)$ as the quantum 
%(field) variable. Instead, we treat the coordinate X as the {\it quantum statistical} 
%variable  using the extra coordinate $\tau$ as the Euclidean time. See Sec.\ref{Qrole} furthermore. 
%}
We impose the periodicity (period: $\beta$) in the direction of the extra dimension $\tau$ (\ref{oneHO2}). 
On a path $\{ x(\tau),\ 0\leq \tau\leq \be\}$, 
the {\it induced} metric is given by
%*** GQSS2 %%%%%%%%%%%%%%%%
\bea
ds^2=(\xdot^2+2V(x))d\tau^2\com\q 0\leq\tau\leq\be
%\left[*********=(1+\frac{\om^2x^2}{\xdot^2})dx^2*******\right]
\pr
\label{GQSS2}
\eea 
%%%%%%%%%%%%%%%%%%%%%%%%%%%
Hence the {\it length} L of the path $x(\tau)$ is given by 
%*** GQSS3%%%%%%%%%%%%%%%%
\bea
L=\int ds=\int_0^\beta\sqrt{\xdot^2+2V(x)}d\tau
\pr
\label{GQSS3}
\eea 
%%%%%%%%%%%%%%%%%%%%%%%%%%%
Taking the half of the length ($\half L$) as the Hamiltonian, 
we get the free energy $F$ as 
%*** GQSS4%%%%%%%%%%%%%%%%
\bea
\e^{-\be F}=\int_{-\infty}^{\infty}d\rho
\int_{\begin{array}{c}x(0)=\rho\\x(\be)=\rho\end{array}}
\prod_\tau\Dcal x(\tau)\exp \left[-\half\int_0^\beta\sqrt{\xdot^2+2V(x)}d\tau\right]
\pr
\label{GQSS4}
\eea 
%%%%%%%%%%%%%%%%%%%%%%%%%%%

\subsection{Standard Type\label{GQSSn}}
%***label***{GQSSn}\nl
We start with the following line element. 
%*** GQSS5%%%%%%%%%%%%%%%%
\bea
ds^2=\frac{1}{d\tau^{2}}(dX^2)^2+4V(X)^2d\tau^2+4V(X)dX^2 
=\frac{1}{d\tau^{2}}\left(  dX^2+2V(X)d\tau^2\right)^2 
\com
\label{GQSS5}
\eea 
%%%%%%%%%%%%%%%%%%%%%%%%%%%]
where we put the condition (\ref{oneHO7}) on the infinitesimal quantities, $d\tau^2$ and $dX^2$,  
in order to keep all terms in the same order. 
The 2D bulk space do {\it not} have 2D metric. 
We impose the periodicity (period: $\be$) (\ref{oneHO8}).
On a path $\{x(\tau),\ 0\leq \tau\leq \be\}$, we have  
the {\it induced} metric: 
%*** GQSS6 %%%%%%%%%%%%%%%%
\bea
ds^2=(\xdot^2+2V(x))^2d\tau^2
%\left[****=(\xdot+\frac{\om^2x^2}{\xdot})^2 dx^2****\right]
\pr
\label{GQSS6}
\eea 
%%%%%%%%%%%%%%%%%%%%%%%%%%%
The {\it length} L is given by 
%*** GQSS7 %%%%%%%%%%%%%%%%
\bea
L[x(\tau)]=\int ds=\int_0^\beta(\xdot^2+2V(x))d\tau
\pr
\label{GQSS7}
\eea 
%%%%%%%%%%%%%%%%%%%%%%%%%%%
Taking $\half L$ as the Hamiltonian, 
the free energy $F$ is given by 
%*** GQSS8%%%%%%%%%%%%%%%%
\bea
\e^{-\be F}=\int_{-\infty}^{\infty}d\rho
\int_{\begin{array}{c}x(0)=\rho\\x(\be)=\rho\end{array}}
\prod_\tau\Dcal x(\tau)\exp \left[-\half\int_0^\beta(\xdot^2+2V(x)) d\tau\right]
\com
\label{GQSS8}
\eea 
%%%%%%%%%%%%%%%%%%%%%%%%%%%
where the path-integral is done for all possible paths with the indicated b.c..  
This is exactly the free energy of the quantum statistical system of 
one variable $x$ in the general potential V($x$).

%%%%%%%%%%%%%%%%%%%%%%%%%%%%  Sec.4  %%%%%%%%%%%%%%%%%%%%%%%%%%%%%%%%%
%%%%                                                            %%%%%%
%%%%  Quantum Statistical System of N Harmonic Oscillators and  %%%%%%
%%%%                          Nonlinear Model             %%%%%%
%%%%                                                            %%%%%%
%%%%%%%%%%%%%%%%%%%%%%%%%%%%%%%%%%%%%%%%%%%%%%%%%%%%%%%%%%%%%%%%%%%%%%
\section{Quantum Statistical System of N Harmonic Oscillators and O(N) Nonlinear Model\label{NHO}}
%***label***{NHO}

\subsection{'Dirac' Type of N Harmonic Oscillators and 
O(N) nonlinear system\label{NHOsup}}
                             %%%   <Fig.4   %%%
\begin{figure}[h]
\includegraphics[width=14pc]{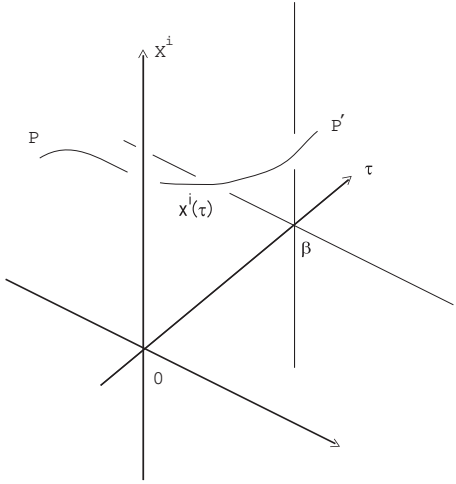}\hspace{2pc}%
\begin{minipage}[b]{14pc}\caption{\label{PathLine}
A path of line $\{x^i(\tau)|i=1,2,\cdots,N\}$ in N(=2)+1 dim space. 
It starts at P=$(\rho_1,\rho_2,\cdots,\rho_N,0)$ and 
ends at P$'$=$(\rho_1',\rho_2',\cdots,\rho_N',\be)$.
%***PathLine.eps\
                                  }
\end{minipage}
\end{figure}
                              %%%   Fig.4>  %%%
%***label***{NHOsup}\nl
Let us consider N+1 dim Euclidean space $(X^i,\tau), i=1,2,\cdots ,N$ described by the following metric. 
%*** NHO1%%%%%%%%%%%%%%%%
\bea
ds^2=\sum_{i=1}^N(dX^i)^2+\om^2 d\tau^2\sum_{i=1}^N(X^i)^2=
\sum_{i=1}^N(dX^i)^2+\om^2 r^2 d\tau^2=
G_{AB}dX^A dX^B\com\nn
A,B=1,2,\cdots,N,N+1;\q X^{N+1}\equiv \tau\com\q \nn
(G_{AB})=\mbox{diag}(1,1,\cdots,1,\om^2r^2)\com\q
r^2\equiv \sum_{i=1}^N(X^i)^2\pr
\label{NHO1}
\eea 
%%%%%%%%%%%%%%%%%%%%%%%%%%%
(Subsec.\ref{1HOsup} is the $N=1$ case. ) 
The Ricci tensor and the scalar curvature are, for N=2, given by
\footnote{
All curvature calculation in this work is checked by 
the algebraic calculation soft "Maxima"\cite{Maxima}. 
}
%*** NHO1b%%%%%%%%%%%%%%%%
\bea
ds^2=dx^2+dy^2+\om^2(x^2+y^2)d\tau^2\com\nn
\left(R_{AB}\right)=\frac{1}{(r^2)^2}
\left(
\begin{array}{ccc}
y^2 & -xy & 0 \\
-yx   & x^2 & 0\\
0    &   0  & \om^2(r^2)^2
\end{array}
\right)
\com\q 
R=\frac{2}{r^2}>0\com\q r^2=x^2+y^2\com\nn
\sqrt{G}=\om\sqrt{x^2+y^2}\com\q
\sqrt{G} R=\frac{2\om}{\sqrt{x^2+y^2}}
\label{NHO1b}
\eea 
%%%%%%%%%%%%%%%%%%%%%%%%%%%
where $(X^1,X^2,X^3)=(x,y,\tau)$ is taken. 
%(See eq.(\ref{GenNHO4}) in App.A, for the general N case using the general potential.)

We impose the periodicity (\ref{oneHO2})(period: $\be$), 
%*** NHO2%%%%%%%%%%%%%%%%
%\bea
%\tau\ra\tau+\be
%\pr
%\label{NHO2}
%\eea 
%%%%%%%%%%%%%%%%%%%%%%%%%%%
and take a path $\{X^i=x^i(\tau)|\ 0\leq \tau\leq \be,\ i=1,2,\cdots,N\}$(See Fig.\ref{PathLine}). 
The {\it induced} metric on the {\it line} is given by
%*** NHO3%%%%%%%%%%%%%%%%
\bea
X^i=x^i(\tau)\com\q 
dX^i=\xdot^i d\tau\com\q \xdot^i\equiv\frac{dx^i}{d\tau}\com\q 
0\leq\tau\leq\be\com
\nn
i=1,2,\cdots,N\com\q
ds^2=\sum_{i=1}^{N}(({\xdot}^i)^2+\om^2(x^i)^2)d\tau^2
\pr
\label{NHO3}
\eea 
%%%%%%%%%%%%%%%%%%%%%%%%%%% 
Then the {\it length} L of the path $\{ x^i(\tau)\}$ is given by 
%*** NHO4%%%%%%%%%%%%%%%%
\bea
L=\int ds=\int_0^\beta\sqrt{ \sum_{i=1}^N((\xdot^i)^2+\om^2{(x^i)}^2) }~d\tau 
\pr
\label{NHO4}
\eea 
%%%%%%%%%%%%%%%%%%%%%%%%%%%
We take the half of the length ($\half L$) as the system Hamiltonian({\it minimal length principle}). 
Then the free energy $F$ of the system is given by 
%*** NHO5%%%%%%%%%%%%%%%%
\bea
\e^{-\be F}=(\prod_i\int_{-\infty}^{\infty}d\rho_i)
\int_{\begin{array}{c}x^i(0)=\rho_i\\x^i(\be)=\rho_i\end{array}}
\prod_{\tau,i}\Dcal x^i(\tau)\exp \left[-\half\int_0^\beta\sqrt{    \sum_{i=1}^N((\xdot^i)^2+\om^2{x^i}^2)}d\tau\right]
\com
\label{NHO5}
\eea 
%%%%%%%%%%%%%%%%%%%%%%%%%%%
where the path-integral is done for all possible paths $\{x^i(\tau);i=1,2,\cdots N\}$ with the indicated b.c.. 
We can regard this as the free energy for 
a variation ('Dirac' type) of the N harmonic oscillators's. 
(See next subsection for the ordinary type of the N harmonic oscillators.)
                             %%%   <Fig.5   %%%
\begin{figure}[h]
\includegraphics[width=14pc]{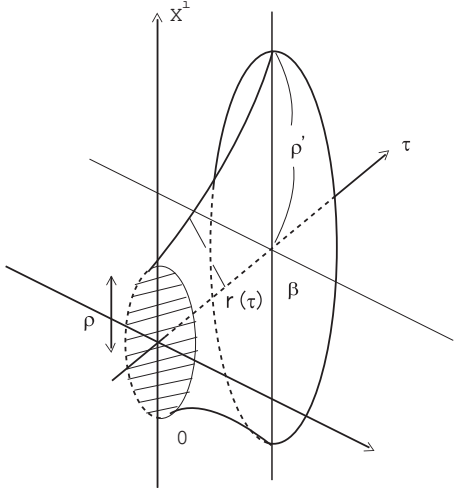}\hspace{2pc}%
\begin{minipage}[b]{14pc}\caption{\label{PathHySurf}
A path of hyper-surface. 
N(=2) dim hypersurface in N+1 dim space $(X^1,X^2,\cdots,X^N,\tau)$. 
S$^{N-1}$ radius $r(\tau)$ starts with $r(0)=\rho$ and ends with $r(\be)=\rho'$. 
We take this configuration as a path in the path integral 
(\ref{NHO9}) and (\ref{NHO18}). This is a closed-string configuration. 
%***PathHySurf.eps\
                                  }
\end{minipage}
\end{figure}
                              %%%   Fig.5>  %%%

Instead of the length $L$, we can take another geometric quantity. Let us consider 
the following N dim {\it hypersurface} in N+1 dim space (a closed-string configuration). 
See Fig.\ref{PathHySurf} for the N=2 case. 
%*** NHO6%%%%%%%%%%%%%%%%
\bea
\sum_{i=1}^N(X^i)^2=r^2(\tau)\com\q
\sum_{i=1}^NX^idX^i=r\rdot d\tau\com\q 0\leq\tau\leq \be
\pr
\label{NHO6}
\eea 
%%%%%%%%%%%%%%%%%%%%%%%%%%%
The form of 
$r(\tau)$ describes a path (N dimensional hypersurface in the bulk) which is {\it isotropic} in 
the 'brane' at $\tau$ (the N dim space 'perpendicularly' standing at $\tau$ of the extra axis, 
not the hypersurface ). 
The {\it induced} metric on the N dim hypersurface is given by
%*** NHO7%%%%%%%%%%%%%%%%
\bea
ds^2=\sum_{i,j}(\del_{ij}+\frac{\om^2}{\rdot^2}x^ix^j)dx^idx^j\equiv
\sum_{i,j}g_{ij}dx^idx^j\com\nn
g_{ij}=\del_{ij}+\frac{\om^2}{\rdot^2}x^ix^j\com\q r^2=\sum_{i=1}^{N}(x^i)^2\com\q
\det(g_{ij})=1+\frac{\om^2r^2}{\rdot^2}
\pr
\label{NHO7}
\eea 
%%%%%%%%%%%%%%%%%%%%%%%%%%%
This is the metric of a O(N) 
nonlinear system and is  
the one dimensional {\it nonlinear sigma model} as the field theory. 
( The standard model (2 dim nonlinear sigma model) 
has often been used so far in order to show the {\it renormalization group} behavior of 
various systems. The background (effective action) formulation of the string 
theory heavily relies on the model.) 
%
%*cite{4D YM to analogy, be-func, asymp free}, 
%*cite{Polyakov $\be_{ij}$-paper, Wilson RG in the coupling space, 
%Friedan 2D nonlinear sigma model 85}. 
Then the {\it area} of the N dim hypersurface is given by 
%*** NHO8%%%%%%%%%%%%%%%%
\bea
A_N=\int\sqrt{\det g_{ij}}~d^Nx=\frac{N\pi^{N/2}}{\Ga(\frac{N}{2}+1)}\int\sqrt{\rdot^2+\om^2 r^2}r^{N-1}d\tau
\pr
\label{NHO8}
\eea 
%%%%%%%%%%%%%%%%%%%%%%%%%%%
When we take $\half A_N$ as the Hamiltonian ({\it minimal area principle}), 
the free energy $F$ is given by 
%*** NHO9%%%%%%%%%%%%%%%%
\bea
\e^{-\be F}=\int_{0}^{\infty}d\rho
\int_{\begin{array}{c}r(0)=\rho\\r(\be)=\rho\end{array}}
\prod_{\tau,i}\Dcal x^i(\tau)\exp \left[
-\half\frac{N\pi^{N/2}}{\Ga(\frac{N}{2}+1)}
\int\sqrt{\rdot^2+\om^2 r^2}r^{N-1}d\tau
                                  \right]
\pr
\label{NHO9}
\eea 
%%%%%%%%%%%%%%%%%%%%%%%%%%%
We should compare this result ($N=4$) with the proposed 5D Casimir energy for the {\it flat} 
geometry (\ref{intro2}). 
The component $\sqrt{\rdot^2+\om^2r^2}$ in the integrand of (\ref{NHO9}) 
is replaced by $\sqrt{\rdot^2+1}$ in (\ref{intro2}).  

We recognize, if we start with 
%*** NHO9c%%%%%%%%%%%%%%%%
\bea
ds^2=\sum_{i=1}^{N}(dX^i)^2+d\tau^2\q (\mbox{N+1 dim Euclidean flat})\com
\label{NHO9c}
\eea 
%%%%%%%%%%%%%%%%%%%%%%%%%%%
instead of (\ref{NHO1}), the integration measure becomes {\it exactly} the same as (\ref{intro2}).

\subsection{Standard Type of N Harmonic Oscillators\label{NHOnrm}}
%***label***{NHOnrm}\nl
Now we consider another type of N+1 dim Euclidean space $(X^i,\tau);\ i=1,2,\cdots N$ described by the following
line element. 
%*** NHO9b%%%%%%%%%%%%%%%%
\bea
ds^2=d\tau^{-2}\{\sum_{i=1}^N(dX^i)^2\}^2+\om^4\{\sum_{i=1}^N(X^i)^2\}^2d\tau^2
+2\om^2\{\sum_{i=1}^N(X^i)^2\} \{\sum_{j=1}^N(dX^j)^2\}  \nn
=\frac{1}{d\tau^{2}}\{ 
\sum_{i=1}^N(dX^i)^2+\om^2r^2d\tau^2
                     \}^2  
\com\q r^2=\sum_{i=1}^N(X^i)^2\com
\label{NHO9b}
\eea 
%%%%%%%%%%%%%%%%%%%%%%%%%%%]
with the condition:  
%*** NHO10%%%%%%%%%%%%%%%%
\bea
\mbox{[Line Element Regularity Condition]}\ :\q\q\q\q\q\q\nn  
d\tau^2\sim O(\ep^2)\com\q (dX^i)^2\sim O(\ep^2)\com\q
\frac{1}{d\tau^2}\{\sum_{i=1}^N(dX^i)^2\}\sim O(1)
\ ,\ 
\label{NHO10}
\eea 
%%%%%%%%%%%%%%%%%%%%%%%%%%%
in order to keep all terms of (\ref{NHO9b}) in the order of $\ep^2$. 
\footnote{
As in (\ref{oneHO7}), this condition restricts the trajectory configuration 
(\ref{NHO12}) only to {\it smooth} hyper-surfaces in the (N+1)-dim space. 
}
Again we 
note that, in the above case, we do {\it not} have N+1 dim (bulk) metric. 
We impose the periodicity (\ref{oneHO2}): (period: $\be$).
%*** NHO11%%%%%%%%%%%%%%%%
%\bea
%\tau\ra\tau+\be
%\pr
%\label{NHO11}
%\eea 
%%%%%%%%%%%%%%%%%%%%%%%%%%%

Here we take a path of Fig.\ref{PathLine}:\ $\{x^i(\tau)|\ 0\leq \tau\leq \be, i=1,2,\cdots,N\}$ and 
the {\it induced} metric on the path is given by
%*** NHO12%%%%%%%%%%%%%%%%
\bea
X^i=x^i(\tau)\com\q dX^i=\xdot^i d\tau\com\q \xdot^i\equiv\frac{dx^i}{d\tau}\com\q 
0\leq\tau\leq\be\com\nn
ds^2=[\sum_{i=1}^N((\xdot^i)^2+\om^2(x^i)^2)]^2 d\tau^2
\pr
\label{NHO12}
\eea 
%%%%%%%%%%%%%%%%%%%%%%%%%%%
Then the {\it length} L of the path $\{x^i(\tau)\}$ is given by 
%*** NHO13%%%%%%%%%%%%%%%%
\bea
L[x^i(\tau)]=\int ds=\int_0^\beta\sum_{i=1}^N((\xdot^i)^2+\om^2(x^i)^2)d\tau
\pr
\label{NHO13}
\eea 
%%%%%%%%%%%%%%%%%%%%%%%%%%%
Hence, taking $\half L$ as the Hamiltonian ({\it minimal length principle}
), 
the free energy $F$ of the system is given by 
%*** NHO14%%%%%%%%%%%%%%%%
\bea
\e^{-\be F}=\left( \prod_i\int_{-\infty}^{\infty}d\rho_i \right)
\int_{\begin{array}{c}x^i(0)=\rho_i\\x^i(\be)=\rho_i\end{array}}
\prod_{i,\tau}\Dcal x^i(\tau)\exp \left[-\half\int_0^\beta\sum_{i=1}^N((\xdot^i)^2+\om^2(x^i)^2)d\tau
                                  \right]
,
\label{NHO14}
\eea 
%%%%%%%%%%%%%%%%%%%%%%%%%%%
where the path-integral is done for all possible paths with the indicated b.c.. This 
is {\it exactly} the free energy 
of N harmonic oscillators. 

We note again the condition (\ref{NHO10}) is necessary for the {\it elastic} 
view to the hyper-surfaces. 

\subsection{Middle type of O(N) nonlinear system\label{NBG}}
%***label***{NBG}\nl
Instead of (\ref{NHO9b}), we can start from a slightly modified metric.   
%*** NHO15%%%%%%%%%%%%%%%%
\bea
ds^2=\om^4\{\sum_{i=1}^N(X^i)^2\}^2d\tau^2
+2\om^2\ka\{\sum_{i=1}^N(X^i)^2\} \{\sum_{j=1}^N(dX^j)^2\}  \nn
=\om^2r^2\left( \om^2r^2d\tau^2+2\ka\sum_{j=1}^N(dX^j)^2\right)\com\q r^2=\sum_{i=1}^N(X^i)^2
\pr
\label{NHO15}
\eea 
%%%%%%%%%%%%%%%%%%%%%%%%%%%
We drop the first term of (\ref{NHO9b}), and add 
a free (real) parameter $\ka$ in the third one. 
We stress that, in this case, we need {\it not} the condition of (\ref{NHO10}). 
The line element is the ordinary type and 
we have the bulk metric $G_{AB}$ in this case. 
The Ricci tensor and the scalar curvature, for $N=2$, are given by 
%*** NHO15b%%%%%%%%%%%%%%%%
\bea
ds^2=\om^4(x^2+y^2)^2d\tau^2+2\om^2\ka (x^2+y^2)(dx^2+dy^2)\com\nn
\left(R_{AB}\right)=\frac{1}{(r^2)^2}
\left(
\begin{array}{ccc}
4y^2 & -4xy & 0 \\
-4yx   & 4x^2 & 0\\
0    &   0  & \frac{2\om^2}{\ka}(r^2)^2
\end{array}
\right)
\com\q 
R=\frac{4}{\ka\om^2(r^2)^2}\com\q r^2=x^2+y^2\com\nn
\sqrt{G}=2\om^4 |\ka| r^4\com\q \sqrt{G}R=8\om^2\cdot\mbox{sign}(\ka)\com
\label{NHO15b}
\eea 
%%%%%%%%%%%%%%%%%%%%%%%%%%%
where $(X^1,X^2,X^3)=(x,y,\tau)$ and sign$(\ka)$ is the sign of $\ka$.
\footnote{
$R>0 \q\mbox{for}\q \ka>0\com\q R<0 \q\mbox{for}\q \ka<0$. 
} 
%(See eq.(\ref{GenNHO15b}) in App.A for general potential.) 
We consider the N dim hypersurface (\ref{NHO6}), or Fig.\ref{PathHySurf}, 
and the {\it induced} metric on it is 
given by 
%*** NHO16%%%%%%%%%%%%%%%%
\bea
ds^2=\sum_{i,j=1}^N 2\om^2r^2(\ka\del_{ij}+\half\frac{\om^2}{\rdot^2}x^ix^j)dx^idx^j\equiv
\sum_{i,j}g_{ij}dx^idx^j
\pr
\label{NHO16}
\eea 
%%%%%%%%%%%%%%%%%%%%%%%%%%%
Then the {\it area} of this hypersurface is given by 
%*** NHO17%%%%%%%%%%%%%%%%
\bea
A_N=\int\sqrt{\det g_{ij}}~d^Nx=
\frac{(2\pi\om^2|\ka|)^{N/2}}{\Ga(\frac{N}{2}+1)}
\int_0^\be r^N\sqrt{\rdot^2+\frac{r^2\om^2}{2|\ka|}}~r^{N-1}d\tau
\pr
\label{NHO17}
\eea 
%%%%%%%%%%%%%%%%%%%%%%%%%%%
Taking $\half A_N$ as the Hamiltonian ({\it minimal area principle}), 
the free energy, $F$, is given by 
%*** NHO18%%%%%%%%%%%%%%%%
\bea
\e^{-\be F}=\int_{0}^{\infty}d\rho
\int_{\begin{array}{c}r(0)=\rho\\r(\be)=\rho\end{array}}
\prod_{\tau,i}\Dcal x^i(\tau)\exp \left[
-\half 
\frac{(2\pi\om^2|\ka|)^{N/2}}{\Ga(\frac{N}{2}+1)}
\int_0^\be r^N\sqrt{\rdot^2+\frac{r^2\om^2}{2|\ka|}}~r^{N-1}d\tau
                                  \right]
\pr
\label{NHO18}
\eea 
%%%%%%%%%%%%%%%%%%%%%%%%%%%
We should compare this result (N=4,$\ka$=1/2) with the proposed 5D Casimir energy for 
the {\it warped} geometry (\ref{intro2}). They are similar 
(
$(\om r)^4\sqrt{\rdot^2+r^2\om^2}$ of (\ref{NHO18}) is replaced by 
$(1/\om z)^4\sqrt{{r'}^2+1}$ of (\ref{intro2}). 
).

\subsection{Modified type of O(N) nonlinear system\label{Mod}}
%***label***{Mod}\nl
Instead of (\ref{NHO15}), we take the following type of metric.   
%*** NHO19b %%%%%%%%%%%%%%%%
\bea
ds^2=
W(\tau)\left( V(r)d\tau^2+\sum_{j=1}^N(dX^j)^2\right)\com\q r^2=\sum_{i=1}^N(X^i)^2
\pr
\label{NHO19b}
\eea 
%%%%%%%%%%%%%%%%%%%%%%%%%%%
Especially, if we start with $W(\tau)=1/\tau^2, V(r)=1$, 
%*** NHO19%%%%%%%%%%%%%%%%
\bea
\mbox{Euclidean }(\mbox{AdS})_{N+1}:\q ds^2=\frac{1}{\tau^2}\{ d\tau^2+\sum_{j=1}^N(dX^j)^2\}  
\com
\label{NHO19}
\eea 
%%%%%%%%%%%%%%%%%%%%%%%%%%%
we recognize the integration measure {\it exactly} becomes 
the same as that in (\ref{intro2}):  
$\tau^{-N}\sqrt{\rdot^2+1}r^{N-1}d\tau$. 
\nl
\nl

The content in this section can be generalized for the {\it general} isotropic potential.

%%%%%%%%%%%%%%%%%%%%%%%%%%%%  Sec.5  %%%%%%%%%%%%%%%%%%%%%%%%%%%%%%%%%
%%%%                                                            %%%%%%
%%%%  Quantum Role of Space-Time Coordinates and                %%%%%%
%%%%  the Matter Fields  - New Treatment of Quantum Gravity     %%%%%%
%%%%                                                            %%%%%%
%%%%%%%%%%%%%%%%%%%%%%%%%%%%%%%%%%%%%%%%%%%%%%%%%%%%%%%%%%%%%%%%%%%%%%
\section{Casimir Energy\label{CE}}
%***label***{CE}
%%%%%%%%%%%%%%%%%%%%%%%%%%%
One important application of the present approach is Casimir energy. 
As mentioned in the introduction, it requires carefull regularization 
to define the quantity rigorously. 
%It is a basic physical quantity in the quantum field theory. It is defined 
%as the vacuum (zero temperature) energy of the free part (interaction-independent part). 
In order to deal with divergence properly, the IR and UV regularizations are 
important. The final (finite) result depends on the {\it boundary} parameters only.

\subsection{4D Flat Case (Ordinary Casimir Energy)}
Before explaining the original motivation (the 5 dim Casimir energy), 
we discuss the ordinary case (1+3 dim electromagnetism) and the relation to the present approach. 
We consider the electromagnetism in Minkwski space: 
%*** CE1%%%%%%%%%%%%%%%%
\bea
ds^2=-dt^2+dx^2+dy^2+dz^2
\pr
\label{CE1}
\eea 
%%%%%%%%%%%%%%%%%%%%%%%%%%%
We place 2 perfectly-conducting plates parallel with the separation $2l$ in the x-direction. 
This configuration can be realized by taking the following boundary conditions. 
As for y- and z-directions, we impose the periodicity for the infra-red (IR) regularization.
%*** CE2%%%%%%%%%%%%%%%%
\bea
\mbox{Periodicity\ \ :\q}
x\ra x+2l\com\q
y\ra y+2L\com\q
z\ra z+2L
\com\nn
L\gg l
\com
\label{CE2}
\eea
%%%%%%%%%%%%%%%%%%%%%%%%%%%
where $L$ is the IR-regularization parameter. Then the eigen frequencies of the electromagnetic wave 
and Casimir energy are given by 
%*** CE3%%%%%%%%%%%%%%%%
\bea
\om_{n,m_y,m_z}=\sqrt{(n\frac{\pi}{l})^2+(m_y\frac{\pi}{L})^2+(m_z\frac{\pi}{L})^2}\com\nn
E_{Cas}=2\cdot\sum_{n,m_y,m_z\in \bfZ}\half\om_{n,m_y,m_z}\q \geq 0
\com
\label{CE3}
\eea
%%%%%%%%%%%%%%%%%%%%%%%%%%%
where $\bfZ$ indicates all integers. 
$\half\om_{n,m_y,m_z}$ is the zero-point oscillation energy. 
Introducing the cut-off function $g(x)$ ($= 1$ for $0<x<1$, 0 for otherwise), Casimir 
energy can formally be written as  
%*** CE4%%%%%%%%%%%%%%%%
\bea
E_{Cas}^{\La}=\sum_{n,m_y,m_z\in \bfZ}\om_{n,m_y,m_z}g\left(\frac{\om_{n,m_y,m_z}}{\La}\right)
\q\geq 0
\pr
\label{CE4}
\eea
%%%%%%%%%%%%%%%%%%%%%%%%%%%
We take the continuum limit $L\ra\infty,\ L\ll l\ra\infty$. 
%*** CE5%%%%%%%%%%%%%%%%
\bea
E_{Cas}^{\La 0}=\int_{-\infty}^{\infty}\int_{-\infty}^{\infty}\frac{dk_ydk_z}{(\frac{\pi}{L})^2}
\int_{-\infty}^{\infty}\frac{dk_x}{\frac{\pi}{l}}\sqrt{k_x^2+k_y^2+k_z^2}g(\frac{k}{\La})\nn
={\int_{-\infty}^{\infty}\int_{-\infty}^{\infty}\int_{-\infty}^{\infty}}_{|k|\leq\La}
\frac{dk_xdk_ydk_z}{(\frac{\pi}{L})^2\frac{\pi}{l}}
\sqrt{k_x^2+k_y^2+k_z^2}\q\geq 0
\pr
\label{CE5}
\eea
%%%%%%%%%%%%%%%%%%%%%%%%%%%
Note that $E_{Cas}$, $E_{Cas}^\La$ and $E_{Cas}^{\La 0}$ are all positive-definite. 

In a familiar way, regarding $E_{Cas}^{\La 0}$ as the origin of the energy scale, we consider 
the quantity $u=(E_{Cas}^{\La}-E_{Cas}^{\La 0})/(2L)^2$ 
as the physical Casimir energy and evaluate it with the help 
of the Euler-MacLaurin formula as $u=(\pi^2/(2l)^3)~(B_4/4!)=-(\pi^2/720)(1/(2l)^3)<0$. 
\footnote{
$B_4=-1/30$ is the 4-th Bernoulli number. 
}
The final result is negative. In the present analysis we take a new regularization. 

\begin{figure}[h]
\begin{minipage}{14pc}
\includegraphics[width=14pc]{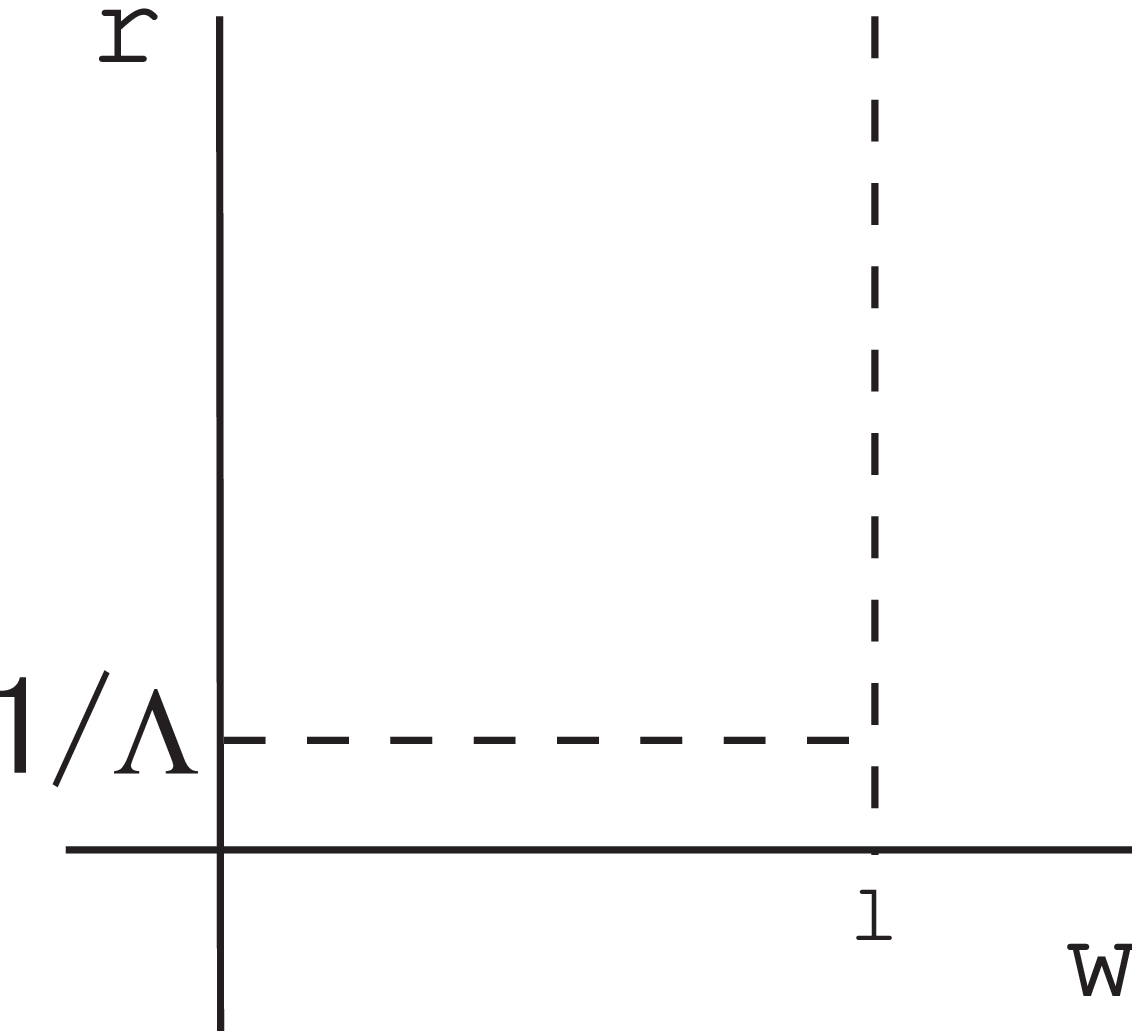}
\caption{\label{2DregionI9}The integral region of (\ref{CE6}). 
%***2DregionI9.eps
}
\end{minipage}\hspace{2pc}%
\begin{minipage}{14pc}
\includegraphics[width=14pc]{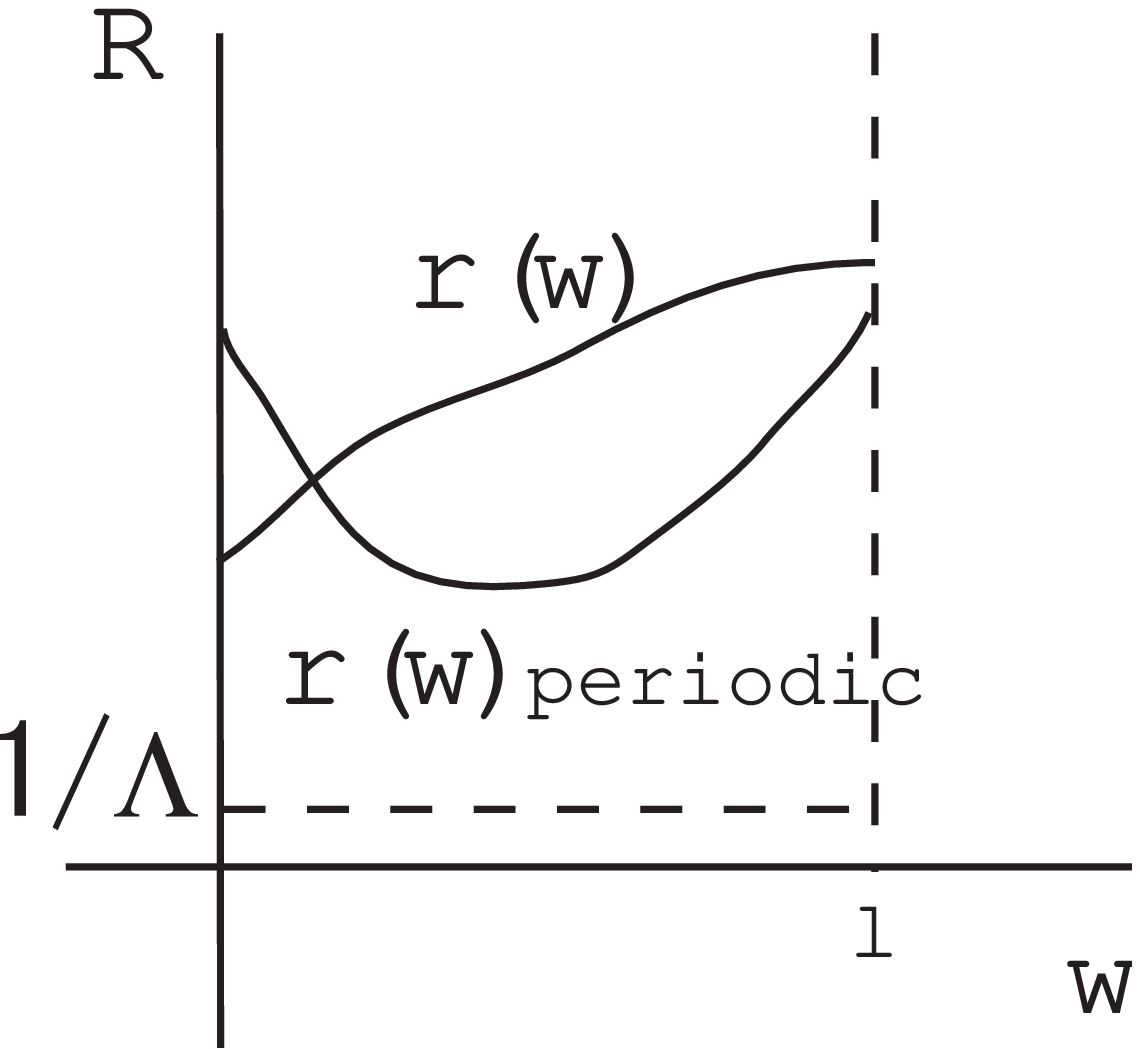}
\caption{\label{2DpathI9}A general path $r(w)$ of (\ref{CE7}) and a periodic path $r(w)$ of (\ref{CE8}). 
%***2DpathI9.eps\
         }
\end{minipage} 
\end{figure}

First we re-express $E_{Cas}^{\La 0}$ using a simple identity : 
$l=\int_0^ldw$ ($w$: a regularization or 'extra' axis). 
%*** CE6%%%%%%%%%%%%%%%%
\bea
E_{Cas}^{\La 0}/(2L)^2
=\frac{1}{2^2\pi^3}\int_0^ldw\int_{k\leq\La}P(k)2\pi k^2dk\nn
=\frac{1}{2^2\pi^3}\int_0^ldw (-1)\int_{r\geq\La^{-1}}P(1/r)(-1)2\pi r^{-4}dr\pr\nn
P(k)\equiv k\com\q r\equiv \frac{1}{k}
\com
\label{CE6}
\eea
%%%%%%%%%%%%%%%%%%%%%%%%%%%
where the integration variable changes from the momentum
($k$) 
%($k=\sqrt{{k_x}^2+{k_y}^2+{k_z}^2}$) 
to the coordinate 
($r=\sqrt{x^2+y^2+z^2}$). The integration region in ($R,w$)-space is the infinite rectangular 
shown in Fig.\ref{2DregionI9}. 
                             %%%   <Fig.5   %%%
%\begin{figure}[h]
%\includegraphics[width=14pc]{2DregionI9.eps}\hspace{2pc}%
%\begin{minipage}[b]{14pc}\caption{\label{2DregionI9}
%The integral region of (\ref{CE6}). 
%%***2DregionI9.eps\
%                                  }
%\end{minipage}
%\end{figure}
                              %%%   Fig.5>  %%%

We regularize the above expression using the path-integral as 
%*** CE7%%%%%%%%%%%%%%%%
\bea
{E_{Cas}^{\Wcal}}'/(2L)^2
=\frac{1}{2^2\pi^3}(2\pi)\int_{\mbox{all paths $r(w)$}}\nn
\prod_w\Dcal r(w) \left[ \int dw'P(\frac{1}{r(w')}) r(w')^{-4}
                 \right]
\exp\left\{ -\Wcal[r(w)]\right\}
\com
\label{CE7}
\eea
%%%%%%%%%%%%%%%%%%%%%%%%%%%
where the integral is over all paths $r(w)$ which are defined between $0\leq w\leq l$ 
and whose value is above $\La^{-1}$, 
as shown in Fig.\ref{2DpathI9}. 
                             %%%   <Fig.6   %%%
%\begin{figure}[h]
%\includegraphics[width=14pc]{2DpathI9.eps}\hspace{2pc}%
%\begin{minipage}[b]{14pc}\caption{\label{2DpathI9}
%A general path $r(w)$ of (\ref{CE7}) and a periodic path $r(w)$ of (\ref{CE8}). 
%%***2DpathI9.eps\
%                                  }
%\end{minipage}
%\end{figure}
                              %%%   Fig.6>  %%%

$\Wcal[r(w)]$ is some damping functional explained in the next paragraph. 
The case $\Wcal[r(w)]=0$ corresponds 
to (\ref{CE6}). The slightly-more-restrictve regularization is 
%*** CE8%%%%%%%%%%%%%%%%
\bea
E_{Cas}^{\Wcal}/(2L)^2
=\frac{1}{2^2\pi^3}(2\pi)\int_{\La^{-1}}^\infty d\rho\int_{r(0)=r(l)=\rho}\nn
\prod_w\Dcal r(w) \left[ \int dw'P(\frac{1}{r(w')}) r(w')^{-4}
                  \right]
\exp\left\{ -\Wcal[r(w)]\right\}\q \geq 0
\com
\label{CE8}
\eea
%%%%%%%%%%%%%%%%%%%%%%%%%%%
where the integral is over all {\it periodic} paths. Note that the above regularization keep 
the positive-definite property. 

Hence the present regularization mainly defined by the choice of $\Wcal[r(w)]$. 
In order to specify it, we introduce the following metric in ($R,w$)-space. 
%*** CE9%%%%%%%%%%%%%%%%
\bea
ds^2=\frac{1}{dw^2}(dR^2+\Omega^2R^2dw^2)^2
\pr
\label{CE9}
\eea
%%%%%%%%%%%%%%%%%%%%%%%%%%%
This is the same as that in Sec.2.2. 
On a path $R=r(w)$, the induced metric and the length $L$ is given as follows. As the damping 
functional $\Wcal[r(w)]$, we take the length $L$. 
%*** CE10%%%%%%%%%%%%%%%%
\bea
ds^2=dw^2({r'}^2+\Omega^2r^2)^2\com\q r'\equiv \frac{dr}{dw}\com\nn
L=\int ds=\int ({r'}^2+\Omega^2r^2)dw\com\q
\Wcal[r(w)]\equiv \frac{1}{2\al}L
=\frac{1}{2\al}\int ({r'}^2+\Omega^2r^2)dw
\pr
\label{CE10}
\eea
%%%%%%%%%%%%%%%%%%%%%%%%%%%
The two parameters $\al$ and $\Omega$ are considered as regularization ones. 
The limit $\al\ra\infty$ corresponds to (\ref{CE6}). 

Numerical calculation can evaluate $E_{Cas}^\Wcal$ (\ref{CE8}), and we expect the 
following form\cite{SI0801,SI0812}. 
%*** CE11%%%%%%%%%%%%%%%%
\bea
\frac{E_{Cas}^\Wcal}{(2L)^2}=\frac{a}{l^3}(1-3c\ln~(l\La))
\com
\label{CE11}
\eea
%%%%%%%%%%%%%%%%%%%%%%%%%%%
where $a$ and $c$ are some constants. 
$a$ should be positive because of the positive-definiteness of (\ref{CE8}). 
The present regularization result has, like the ordinary 
renormalizable ones such as the coupling 
in QED, the log-divergence. The divergence can be renormalized into the boundary parameter $l$. 
This means $l$ {\it flows} according to the renormalization group. 
%*** CE12%%%%%%%%%%%%%%%%
\bea
l'=l(1-3c\ln(l\La))^{-\frac{1}{3}}\com\q
\be\equiv \frac{d\ln(l'/l)}{d\ln\La}=c\com\q |c|\ll 1
\com
\label{CE12}
\eea
%%%%%%%%%%%%%%%%%%%%%%%%%%%
where $\be$ is the renormalization group function, and we assume $|c|\ll 1$. 
The sign of $c$ determines whether the length separation increases ($c>0$) or decreases ($c<0$) 
as the measurement resolution becomes finer ($\La$ increases). In terms of the usual terminology, 
attractive case corresponds to $c>0$, and repulsive case to $c<0$.

\subsection{5D Flat and Curved Case}
%Traditional calculation\cite{AC83,SI0801,SI0812} gives 
%the $\La^5$-divergent result for Casimir energy of the above models. 
%The {\it closed} expression of the energy has recently been presented 
%both for the flat\cite{SI0801} and for the warped\cite{SI0812} cases as follows.
Let us mention the space-time quantization (quantum gravity) 
in relation to the motivation of the present work. 
The space-time geometry is specified by the metric tensor field $g_\mn(x)$ 
which appears in the definition of 
the line element $(ds^2)_{4D}=g_\mn(x)dx^\m dx^\n~(\m,\n=0,1,2,3)$. 
One of most important problems of the present theoretical physics is the clarification 
of the {\it quantum role} of the metric (gravitational) field $g_\mn$. We already have 
a long (nearly half century) history of the quantum gravity since Feynman\cite{Fey63} and 
DeWitt \cite{DeW67} pioneered. 
About one decade ago, inspired by the development of the string theory and 
the D-brane theory, a fascinating model of unification of forces 
was proposed. It is a 5 dimensional(dim) model with 
AdS$_5$ geometry and is called "Randall-Sundrum model" or the "warped model"\cite{RS99}. 
This is a representative of the extra dimensional models. 
An important purpose of the present work is to make this 5 dim model 
{\it legitimate} as the {\it quantum field theory}. 

In the warped space-time
%\footnote{
%Anti de Sitter (AdS) space is a maximally-symmetric constant-curvature 
%space-time with the negative cosmological constant. 
%%It is also called "warped" space-time. 
%%Here we consider 5 dimensional 
%%anti de Sitter (AdS$_5$ ) as the curved case. 
%}
, 
the geometry is described as
%*** intro1B%%%%%%%%%%%%%%%%
\bea
\mbox{Warped Metric (y-expression)}\q
ds^2=\e^{-2\om |y|}\eta_\mn dx^\m dx^\n +dy^2\ ,\ 
-l\leq y\leq l\ ,
\label{intro1B}
\eea 
%%%%%%%%%%%%%%%%%%%%%%%%%%%
where 
$
\{\mu,\nu=0,1,2,3\}\ ,\ 
(\eta_\mn)=\mbox{diag}(-1,1,1,1) 
$. The coordinate 
$y$ is called the extra coordinate. The parameter $\om$ is the 5 dim (bulk) scalar curvature. 
$l$ is the size parameter of the extra coordinate. 
We respect the periodicity: $y\ra y+2l$\ , and Z$_2$-parity: $y\change -y$. 
Instead of $y$, another coordinate $z$ is also used. 
\footnote{
%*** intro1C  ****
$z$ is defined by $y$ as
\bea
z=\left\{
\begin{array}{cc}
\frac{1}{\om}\e^{\om y} &  y>0 \\
0    &  y=0 \\
-\frac{1}{\om}\e^{-\om y} &  y<0 
\end{array}
\right.
\nonumber
%\label{intro1C}
\eea
}
%*** intro1x%%%%%%%%%%%%%%%%
\bea
\mbox{Warped Metric (z-expression)}\q
ds^2=\frac{1}{\om^2z^2}(\eta_\mn dx^\m dx^\n+{dz}^2)
=G_{MN}dX^MdX^N \ ,\nn
|z|=\frac{1}{\om}\e^{\om|y|}
\com\q   \frac{1}{\om}<|z|<\frac{1}{T}\com\q
T\equiv \om\e^{-\om l}\com
% G\equiv \det G_{AB}\com
\nn 
R_{MN}=4\om^2G_{MN}
 \com\q R=20\om^2>0
\com\q
\sqrt{-G}=\sqrt{-\det G_{MN}}=\frac{1}{(\om|z|)^5}\com
\label{intro1x}
\eea 
%%%%%%%%%%%%%%%%%%%%%%%%%%%
where 
$
(X^M)\equiv (x^\mu,z)\ ,\ \{M,N=0,1,2,3,5\}
$. 
\footnote{
$T$ is {\it not} a temperature parameter but a IR parameter like $l$ ($T=\om\e^{-\om l}$). 
The temperature appears as $\be^{-1}$. See eq.(\ref{oneHO2}). 
}
The flat (5D Minkowski) limit is obtained by $\om\ra 0$ in the y-expression (\ref{intro1B}). 
%*** intro1b%%%%%%%%%%%%%%%%
\bea
\mbox{Flat Metric}\q
ds^2=\eta_\mn dx^\m dx^\n +dy^2\com\q (X^M)=(x^\m, y)\com\q 
-l\leq y\leq l
\com
\label{intro1b}
\eea 
%%%%%%%%%%%%%%%%%%%%%%%%%%%

\begin{figure}[h]
\includegraphics[width=14pc]{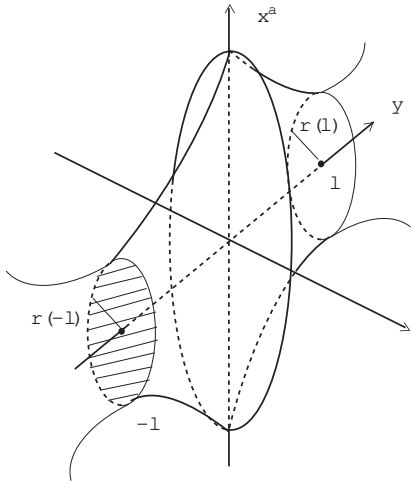}\hspace{2pc}%
\begin{minipage}[b]{14pc}\caption{\label{HySurfF}
N(=2) dim hypersurface in N+1 dim (Euclidean flat) space $(x^1,x^2,\cdots,x^N,y)$. 
Sphere S$^{N-1}$(circles in the figure) at $y$ has the radius $r(y)$. 
%***HySurfF.eps\
                                  }
\end{minipage}
\end{figure}

Traditional calculation\cite{AC83,SI0801,SI0812} gives 
the $\La^5$-divergent result for Casimir energy, on the above geometries, 
of 5D models. 
In the calculation, Casimir energy is expressed as 
the 5D space-momentum integral ($\int d^4p_Edy$ or $\int d^4p_Edz$) of 
some energy (density) function $F(\ptil,y)$ or $F(\ptil,z)$. (See Appendix for detail.) 
In ref.\cite{SI0801,SI0812}, 
we claim the $\La^5$-divergence comes from this 'naive' integration measure 
and should be replaced by some proper measure, based on close numerical 
calculation using some trial integration measures. 
Finally, Casimir energy of 
the free fields (electromagnetism, free scalar theory) 
is {\it proposed} to be replaced by the 
following {\it path-integral}. 
%*** intro2 %%%%%%%%%%%%%%%%
\bea
\mbox{For Flat Geometry\ :}\qqqqq\qqqqq\qqqqq\nn
-\Ecal_{Cas}(l,\La)
=\int_{1/\La}^{l}d\rho\int_{r(0)=r(l)=\rho}
\prod_{a,y}\Dcal x^a(y) \left[ \int dy'F_1(\frac{1}{r(y')},y')
                        \right]
\mbox{\ }\nn
\times\exp\left[ 
-\frac{1}{2\al'}\int_{0}^{l}\sqrt{{r'}^2+1}~r^3 dy
    \right],\ r'=\frac{dr}{dy},\nn
\mbox{For Warped Geometry\ :}\qqqqq\qqqqq\qqqqq\nn
-\Ecal_{Cas}(\om,T,\La) 
=\int_{1/\La}^{1/\m}d\rho\int_{r(1/\om)=r(1/T)=\rho}
\prod_{a,z}\Dcal x^a(z) \left[ \int dz'F_2(\frac{1}{r(z')},z')
                        \right]\nn
\times\exp\left[ 
-\frac{1}{2\al'}\int_{1/\om}^{1/T}\frac{1}{\om^4z^4}\sqrt{{r'}^2+1}~r^3 dz
    \right]\com\q r'=\frac{dr}{dz}\com
\label{intro2}
\eea 
%%%%%%%%%%%%%%%%%%%%%%%%%%%
where 
$r=\sqrt{\sum_{a=1}^{4}(x^a)^2}$. 
\footnote{
The case $\al'\ra\infty$ in (\ref{intro2}) is essentially the traditional 
definition of Casimir energy. 
}
($\{x^a| a=1,2,3,4\}$ is the {\it Euclideanized} coordinates of 
$\{x^\m |~\mu=0,1,2,3\} $, $x^0=ix^4$.)
$F_1$ and $F_2$ are some energy density functions and
will appear in Appendix ((\ref{HK20}) and (\ref{HKA11})). 
$\La$ is the UV-cutoff parameter, 
$\m\equiv\La T/\om$ is the IR-cutoff one and $l$ is the periodicity(IR) one. 
Note that the above expressions of $-\Ecal_{Cas}$ are positive-definite. 
The above path-integrals are over all paths 
of 4 dim {\it hypersurfaces} defined by 
%*** intro3 %%%%%%%%%%%%%%%%
\bea
\mbox{Flat Geometry:}\q\q\sqrt{\sum_{a=1}^{4}(x^a)^2}=r(y)\com\q -l\leq y \leq l\com\nn
\mbox{Warped Geometry:}\q\q\sqrt{\sum_{a=1}^{4}(x^a)^2}=r(z)\com\q \frac{1}{\om}\leq |z| \leq \frac{1}{T}\pr
\label{intro3}
\eea 
%%%%%%%%%%%%%%%%%%%%%%%%%%%
See Fig.\ref{HySurfF} for the case of the N+1 dim space. 
                             %%%   <Fig.1   %%%
                              %%%   Fig.1>  %%%
This is a {\it closed string} configurartion. 
The {\it area} (4D volume) plays the role of {\it Hamiltonian} of the 
quantum statistical system $\{x^a\}$. 
$F_i$ comes from the {\it matter-field} quantization and   
plays a role of the {\it energy 'operator'} in the path-integral over the 
4D hyper-surface $r(y)$ or $r(z)$. 
The {\it string (surface) tension} parameter $1/2\al'$ is introduced.  
%Note that the above result is obtained by taking the {\it new standpoint} that 
%the (bulk) metric field $G_{MN}(X)$ 
%is {\it not} quantized and is 
%treated as a background field. 
%%\footnote{
%%We consider that the form of $G_{MN}(X)$ is given by the 
%%field equation of the 'effective' action which is obtained 
%%after the {\it field} quantization of all {\it matter} fields. 
%%It is a fixed (or background) field in the quantization process of the space-time. 
%%See also Sec.\ref{Qrole}. 
%%}
%Instead we regard the 
%4 dim coordinates $x^a$ as the quantum (statistical) variables, and 
%the extra one, $y$ or $z$, as Euclidean time. 
The new point, compared with the 5D Casimir energy calculation so far\cite{AC83}, 
is the introduction of the 'minimal area' factor 
$\exp (-\frac{1}{2\al'}\mbox{Area})=\exp (-\frac{1}{2\al'}\int\sqrt{\det(g_{ab})}d^4x)$ 
where $g_{ab}$ is the {\it induced} metric on the hyper-surface (\ref{intro3}).  
We have shown, in this paper, 
the above-type {\it path-integral} very naturally appears in many 
quantum-statistical systems when we view them {\it geometrically}. 
The proposed quantities (\ref{intro2}) are shown to be valid. 
%This is the final aim of this paper. 

The proposed expressions (\ref{intro2}) of 5D Casimir energy 
can be evaluated numerically. We confirmed, not using the path-integral 
but using some effective approach (weight-function method), 
that they are given by as follows. 
%*** 5DCas %%%%%%%%%%%%%%%%
\bea
\mbox{Flat\ }:\q\frac{\Ecal_{Cas}(l,\La)}{\La l} =-\frac{a}{l^4}(1-4c\ln(l\La))\ ,\q
a\sim 2.5\com\q c>0\com\q c\sim O(10^{-3})\com\nn
\mbox{Warped\ }:\q\frac{\Ecal_{Cas}(\om,T,\La)}{\La T^{-1}}=-a\om^4(1-4c_1\ln(\frac{\La}{\om}) 
                      -4c_2\ln(\frac{\La}{T}))\ ,\q\nn
a\sim 1.2\ ,\ c_1\sim -0.11 < 0\ ,\ c_2\sim 0.10 > 0\ .
\label{5DCas}
\eea 
%%%%%%%%%%%%%%%%%%%%%%%%%%%
The boundary parameters {\it flow} as
%*** 5DCas2 %%%%%%%%%%%%%%%%
\bea
\mbox{Flat\ }:\q\frac{\Ecal_{Cas}(l,\La)}{\La l} =-\frac{a}{{l'}^4}\com\q
\be=\frac{\pl}{\pl (\ln\La)}\ln\frac{l'}{l}=c>0\com\nn
\mbox{Warped\ }:\q\frac{\Ecal_{Cas}(\om,T,\La)}{\La T^{-1}}=-a{\om'}^4\com\q
\be=\frac{\pl}{\pl (\ln\La)}\ln\frac{\om'}{\om}=-c_1-c_2\pr
\label{5DCas2}
\eea 
%%%%%%%%%%%%%%%%%%%%%%%%%%%
$\be=c>0$ in the flat case means the size of the extra world 
(periodicity $l$) shrinks as the measurement resolution becomes coarse. 
This image fits with the compactification of the extra axis 
in the higher-dimensional unified models.

%%%%%%%%%%%%%%%%%%%%%%%%%%%%  Sec.6  %%%%%%%%%%%%%%%%%%%%%%%%%%%%%%%%%
%%%%                                                            %%%%%%
%%%%  Visco-Elastic System                                 %%%%%%
%%%%                                                            %%%%%%
%%%%%%%%%%%%%%%%%%%%%%%%%%%%%%%%%%%%%%%%%%%%%%%%%%%%%%%%%%%%%%%%%%%%%%
\section{Visco-Elastic System\label{VE}}
%***label***{VE}
The present approach gives a new method for the study of 
the visco-elastic system. Let us explain it using the 
harmonic oscillator with friction. 
%*** ve1 %%%%%%%%%%%%%%%%
\bea
m\xddot=-kx-\eta\xdot\ ,\ k:\ \mbox{spring constant},\ \ \eta:\ \mbox{viscosity}\pr
\label{ve1}
\eea 
%%%%%%%%%%%%%%%%%%%%%%%%%%%
From this we obtain the "energy" at $t=t_0$. 
%*** ve2 %%%%%%%%%%%%%%%%
\bea
\Ecal=
\left(
\half\xdot^2+\frac{{\om_1}^2}{2}x^2
\right)|_{t_0}
=\half\xdot^2+\frac{{\om_1}^2}{2}x^2+\eta'\int^t_{t_0}\left(\frac{dx(\ttil)}{d\ttil}\right)^2d\ttil
\ ,\q{\om_1}^2=\frac{k}{m},\ \eta'=\frac{\eta}{m}.
\label{ve2}
\eea 
%%%%%%%%%%%%%%%%%%%%%%%%%%%
This quantity is conserved (independent of $t$). The hysteresis term in the 
above expression represents the energy from the friction force. We can read 
the line elements in ($X,t$) space for this system. 
%*** ve3 %%%%%%%%%%%%%%%%
\bea
\mbox{'Dirac' type\ :}\q
ds^2=dX^2+dt^2\left( {\om_1}^2X^2+2\eta'\int^t_{t_0}(dX)^2\frac{1}{d\ttil}\right)\com\nn
\mbox{Standard type\ :}\q
ds^2=\frac{1}{dt^2}\left\{ dX^2+dt^2\left({\om_1}^2X^2+2\eta'\int^t_{t_0}dX^2\frac{1}{d\ttil}\right)\right\}^2\nn
\label{ve3}
\eea 
%%%%%%%%%%%%%%%%%%%%%%%%%%%
On a path $X=x(t)$, with the standard type, the {\it length} $L[x(t)]$ and 
the free energy $F$ is given by ($t_0=0$) 
%*** ve4 %%%%%%%%%%%%%%%%
\bea
L[x(t)]=\int ds=
\int^\be_0 dt\{
\xdot^2+{\om_1}^2x^2+2\eta'\int^t_{0}(\frac{dx(\ttil)}{d\ttil})^2d\ttil
       \}                                                        \com\nn
\e^{-\be F(l,\be)}=\int^{l}_{-l}d\rho\int_{x(0)=\rho,x(\be)=\rho}
\Dcal x(t)\e^{-\half L[x(t)]} \com      
\label{ve4}
\eea 
%%%%%%%%%%%%%%%%%%%%%%%%%%%
where $2l$ is the periodicity, with which we impose $X$ periodic. 
%*** ve4a %%%%%%%%%%%%%%%%
\bea
X\ra X+2l
\pr      
\label{ve4a}
\eea 
%%%%%%%%%%%%%%%%%%%%%%%%%%%
The energy $E$, the entropy $S$ and the force $f$ are given by 
%*** ve4b %%%%%%%%%%%%%%%%
\bea
\mbox{Energy}\q E(l,\be)=<\frac{L}{2}>=
\int_{-l}^ld\rho\int\Dcal x(t)\frac{L[x(t)]}{2}\exp\{-\half L[x(t)]\}\com\nn
\mbox{Entropy}\q S(l,\be)=k_B\be (E(l,\be)-F(l,\be))\com\q
\mbox{Force}\q f(l,\be)=-\frac{\pl E(l,\be)}{\pl l}\com
\label{ve4b}
\eea 
%%%%%%%%%%%%%%%%%%%%%%%%%%%
where $k_B$ is Boltzmann's constant. 

The simple model (\ref{ve1}) can be generalized as 
%*** ve4c %%%%%%%%%%%%%%%%
\bea
m\xddot=-\frac{\pl V(x)}{\pl x}-\eta_1\xdot W_1(x)
-\eta_2\xdot^2 W_2(x)\com
\label{ve4c}
\eea 
%%%%%%%%%%%%%%%%%%%%%%%%%%%
where we assume the visco-elastic system is in the relatively 
slow motion. This assumption makes the velocity($\xdot$)-expansion of (\ref{ve4c}) 
valid. $V(x), W_1(x), W_2(x)$ are general functions. There are many choices:\ 
%*** ve4d %%%%%%%%%%%%%%%%
\bea
V(x)=\half kx^2\ (\mbox{spring}),\ gx\ (\mbox{rain-drop}),\ \cdots\nn
W_1(x)=1\com\q x^2-1\ (\mbox{van der Pol eq.}),\ \cdots\nn
W_2(x)=1\com\q x\com\q x^2\com\q x^3\com\q\cdots
\label{ve4d}
\eea 
%%%%%%%%%%%%%%%%%%%%%%%%%%%
The line elements in ($X,t$) space is given by 
%*** ve4e %%%%%%%%%%%%%%%%
\bea
\mbox{'Dirac' type\ :}\qqqqqq\qqqqq\qqqqq\nn
ds^2=m~dX^2+dt^2\left\{ 2V(X)+2\eta_1\int_{t_0}^t(dX)^2\frac{1}{d\ttil}W_1(X)
                             +2\eta_2\int_{t_0}^t(dX)^3\frac{1}{(d\ttil)^2}W_2(X)
                 \right\},\nn
\mbox{Standard type\ :}\qqqqqq\qqqqq\qqqqq\nn
ds^2=\frac{1}{dt^2}\left[ m~dX^2+dt^2\left\{ 2V(X)+2\eta_1\int_{t_0}^t(dX)^2
      \frac{1}{d\ttil}W_1(X)         \right.
                    \right.\nn
                                    \left.
                  \left. 
+2\eta_2\int_{t_0}^t(dX)^3\frac{1}{(d\ttil)^2}W_2(X)
                                     \right\}
                  \right]^2.
\label{ve4e}
\eea 
%%%%%%%%%%%%%%%%%%%%%%%%%%%

\begin{figure}[h]
\includegraphics[width=14pc]{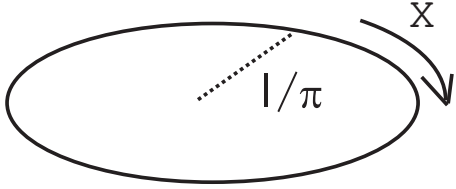}\hspace{2pc}%
\begin{minipage}[b]{14pc}\caption{\label{CircFlowI9}
One dimensional, circling, visco-elastic flow. The radius $l/\pi$ is sufficiently large.  
%***CircFlowI9.eps\
                                  }
\end{minipage}
\end{figure}

The above formulas are valid for the spacially-circling 1-dim 
visco-elastic fluid (Fig.\ref{CircFlowI9}). 
                             %%%   <Fig.7   %%%
                              %%%   Fig.7>  %%%
The global (statistical) physical quantities can be 
expressed by the path-integral in the same way as eq.(\ref{ve4b}). 

For the spacially 3 dim visco-elastic system, we treat   
Navier-Stokes equation (\ref{intro1}). We obtain the following relation. 
%*** ve5 %%%%%%%%%%%%%%%%
\bea
\frac{D}{Dt}\left(\frac{\rho}{2}\vvec^2+P-\rho\xvec\cdot\gvec
            \right) 
=\frac{\pl P}{\pl t}+\eta v^i\Del v^i
\pr
\label{ve5}
\eea 
%%%%%%%%%%%%%%%%%%%%%%%%%%%
From this, we obtain, for the case $\frac{\pl P}{\pl t}=0$, 
%*** ve6 %%%%%%%%%%%%%%%%
\bea
\Ecal=\left(\frac{\rho}{2}\vvec^2+P-\rho\xvec\cdot\gvec
      \right) |_{t_0}
= \frac{\rho}{2}\vvec^2+P-\rho\xvec\cdot\gvec-\eta\int_{t_0}^t v^i\Del v^id\ttil
\pr
\label{ve6}
\eea 
%%%%%%%%%%%%%%%%%%%%%%%%%%%
We can read, from the above result, the following line elements. 
%*** ve7 %%%%%%%%%%%%%%%%
\bea
\mbox{Dirac type\ :}\q 
ds^2=\rho~{dX^i}^{~2}+dt^2\left\{
2P(X)-2\rho X^ig^i-2\eta\int_{t_0}^tdX^i\frac{1}{d\ttil}\Del(dX^i\frac{1}{d\ttil})~d\ttil
                                    \right\},\nn
\mbox{Standard type\ :}\qqqqq\qqqqq\qqqqq\qqqqq\qqqqq\qqqqq \nn
ds^2=\frac{1}{dt^2}\left[
\rho~{dX^i}^{~2}+dt^2\left\{
2P(X)-2\rho X^ig^i-2\eta\int_{t_0}^tdX^i\frac{1}{d\ttil}\Del(dX^i\frac{1}{d\ttil})~d\ttil
                                    \right\}
                    \right]^2
.
\label{ve7}
\eea 
%%%%%%%%%%%%%%%%%%%%%%%%%%%
On a path $X^i=x^i(t)$, with the standard type, the {\it length} $L[\xvec(t)]$ and 
the free energy $F$ are given by ($t_0=0$) 
%*** ve8 %%%%%%%%%%%%%%%%
\bea
L[\xvec(t)]=\int ds=
\int^\be_0 dt\left\{
\rho({\xdot}^i)^2
+2P(\xvec)-2\rho\xvec\cdot\gvec-2\eta\int^t_{0}{\xdot}^i\Del{\xdot}^i~d\ttil
             \right\}                                                        \com\nn
\e^{-\be F(l,\be)}= \int^{l}_{-l}d\rho_1\int^{l}_{-l}d\rho_2\int^{l}_{-l}d\rho_3
\int_{x^i(0)=x^i(\be)=\rho^i}
\Dcal x^i(t)\e^{-\half L[\xvec(t)]} \com      
\label{ve8}
\eea 
%%%%%%%%%%%%%%%%%%%%%%%%%%%
where $2l$ is the periodicity, with which we impose $X^i$ periodic. 
%*** ve9 %%%%%%%%%%%%%%%%
\bea
X^i\ra X^i+2l\q\q (i=1,2,3)
\pr      
\label{ve9}
\eea 
%%%%%%%%%%%%%%%%%%%%%%%%%%%

%%%%%%%%%%%%%%%%%%%%%%%%%%%%  Sec.7  %%%%%%%%%%%%%%%%%%%%%%%%%%%%%%%%%
%%%%                                                            %%%%%%
%%%%  Discussion and Conclusion                                 %%%%%%
%%%%                                                            %%%%%%
%%%%%%%%%%%%%%%%%%%%%%%%%%%%%%%%%%%%%%%%%%%%%%%%%%%%%%%%%%%%%%%%%%%%%%
\section{Discussion and Conclusion\label{conc}}
%***label***{conc}

Inspired by the recent new views on the quantum gravity, we have presented 
a geometrical approach to the general quantum statistical system. 
The idea lies in the introduction of the metric in the space of 
Feynman's path-integral. The length or the area gives the system Hamiltonian. 
In the application to 5D Casimir energy, the inverse temperature (or proper time) 
axis is played by the extra (5th) one. In the 4D case (ordinary Casimir energy), 
the regularization axis plays the role. Casimir force is explained from the 
{\it renormalization group flow}. 
The attractive force or the repulsive one corresponds to 
the positive $\be$-function or the negative one. This geometrical approach also 
gives a new method to examine the dissipative system caused by friction. Taking simple 
visco-elastic models, we elaborate on how to choose the metric and present 
the path-integral expressions 
of the statistical quantities such as energy and entropy.

%We have shown some quantum statistical systems of N variables can be 
%described by the path (line or hypersurface) integral over the N+1 dim Euclidean space 
%with an appropriate Hamiltonian ({\it length} of the line or {\it area} of 
%the hypersurface). The system dynamics is determined by choosing the following 
%two things:\ 
%1)\ With which bulk metric does one start and 2)\ which type of path (line or hypersurface) 
%does one take.  
%In other words, ***************
%the choice 1) specifies the bulk geometry and 
%the choice 2) specifies the embedded geometry of the path. *****************
%This is the {\it geometric view} of the quantum statistical system. The result is applied to 
%Casimir energy of 5 dim models and we show the proposed new definition (\ref{intro2}) is valid. 

\appendix
%%%%%%%%%%%%%%%%%%%%%%%%%%%%  Sec.8  %%%%%%%%%%%%%%%%%%%%%%%%%%%%%%%%%
%%%%                                                            %%%%%%
%%%%  Appendix:\ Traditional Definition of Casimir Energy in    %%%%%%
%%%%              5D Theories                                   %%%%%%
%%%%                                                            %%%%%%
%%%%%%%%%%%%%%%%%%%%%%%%%%%%%%%%%%%%%%%%%%%%%%%%%%%%%%%%%%%%%%%%%%%%%%
\section{
Traditional Definition of Casimir Energy in 5D Theories
\label{app}}
%***label***{app}

The traditional definition of Casimir Energy of the 5D electromagnetic field 
theory is, for the flat case (\ref{intro1b}), 
%*** HK2%%%%%%%%%%%%%%%%
\bea
\e^{-l^4E_{Cas}}=\left.\int\Dcal A \exp\left[
i\intxy(\Lcal^{5D}_{EM}+\Lcal_{gauge})
                                       \right]
                   \right|_{\mbox{Euclid}}\com\hspace{40mm}\nn
\Lcal^{5D}_{EM}[A_M(X)]=-\fourth F_{MN}F^{MN}\ ,\ 
F_{MN}=\pl_MA_N-\pl_NA_M\ ,\ 
\Lcal_{gauge}[A_M(X)]=-\half(\pl_MA^M)^2\ .                   
\label{HK2} 
\eea
%%%%%%%%%%%%%%%%%%%%%%%%%%%%%
The expression of $E_{Cas}$ defined above, is given by 
%*** HK20%%%%%%%%%%%%%%%%
\bea
\hspace{50mm}\mbox{For Flat Geometry (5 dim elctromagnetism) :}\nn
E_{Cas}(l)=\intpL\int_0^ldy (F_f^-(\ptil,y)+4F_f^+(\ptil,y))\com\nn
F_f^\mp(\ptil,y)
=-\int_\ptil^\infty d\ktil\frac{\mp\cosh\ktil(2y-l)+\cosh\ktil l}{2\sinh(\ktil l)}
\pr
\label{HK20}
\eea
%%%%%%%%%%%%%%%%%%%%%%%%%%%%%
The plus-minus symbol, $\mp$, indicates the contribution from 
Z$_2$-parity odd (-) and even (+) components. 
$\ptil$ is the maginitude of 4D momentum $(p_a)=(p_1,p_2,p_3,p_4)$. 
The coincidence with the previous result\cite{AC83} was confirmed\cite{SI0801}. 
As for the warped case (\ref{intro1x}), the traditional definition, for the 5D free scalar theory, is 
given by
%*** KKexp1b%%%%%%%%%%%%%%%%
\bea
\e^{-T^{-4}E_{Cas}}=\left.\int\Dcal\Phi~\exp
\left[
i\int d^5X\sqrt{-G}\Lcal^{5D}_s
\right]              \right|_{\mbox{Euclid}}\nn
=\int\Dcal\Phi(X)\exp\left[
\intfx dz\frac{1}{(\om z)^5}\half\Phi\{
\om^2z^2\pl_a\pl^a\Phi+(\om z)^5 \Lhat_z \Phi
                                       \} 
                     \right]\com\nn
\Lcal^{5D}_s[\Phi(X);X]=-\half \na^M\Phi\na_M\Phi-\half m^2\Phi^2\com\nn
\frac{1}{\om}<|z|<\frac{1}{T}\com\q
\Lhat_z=\frac{d}{dz}\frac{1}{(\om z)^3}\frac{d}{dz}-\frac{m^2}{(\om z)^5}
\com\q (m^2=-4\om^2)
\pr
\label{KKexp1b}
\eea 
%%%%%%%%%%%%%%%%%%%%%%%%%%%
where 
$\Lhat_z$ is the kinetic operator in the extra space (Bessel differential operator). 
Casimir energy $E_{Cas}$ defined in (\ref{KKexp1b}) is explicitly given by 
%*** HKA11%%%%%%%%%%%%%%%%
\bea
\hspace{50mm}\mbox{For Warped Geometry (5 dim Free Scalar, $m^2=-4\om^2$):}\nn
-E^\mp_{Cas}(\om,T)
=\left.\intpE\right|_{\ptil\leq\La}\int_{1/\om}^{1/T}dz~F_w^\mp(\ptil,z)\com 
\q
F_w^\mp(\ptil,z)= \frac{1}{(\om z)^3}\int_{\ptil^2}^\infty\{G_k^\mp (z,z)\}dk^2\com\nn
G_p^\mp(z,z')=\mp\frac{\om^3}{2}z^2{z'}^2
\frac{\{\I_0(\Pla)\K_0(\ptil z)\mp\K_0(\Pla)\I_0(\ptil z)\}  
      \{\I_0(\Tev)\K_0(\ptil z')\mp\K_0(\Tev)\I_0(\ptil z')\}
     }{\I_0(\Tev)\K_0(\Pla)-\K_0(\Tev)\I_0(\Pla)},\nn     
(\Lhat_z-p^2s(z))G^{\mp}_p(z,z')=
\left\{
\begin{array}{ll}
\ep(z)\ep(z')\delh (|z|-|z'|) & \mbox{for\ \ P=}-1 \\
\delh (|z|-|z'|) & \mbox{for\ \ P=}1 
\end{array}
        \right.
\com\q s(z)=\frac{1}{(\om z)^3}\com         
\label{HKA11}
\eea
%%%%%%%%%%%%%%%%%%%%%%%%%%%%%
where $\I_0$ and $\K_0$ are the modified Bessel functions of 0-th order. 
%Both in (\ref{HK20}) and in (\ref{KKexp1b}), Casimir energy is expressed as 
%the 5D space-momentum integral ($\int d^4p_Edy$ or $\int d^4p_Edz$) of 
%energy (density) $F(\ptil,y)$ or $F(\ptil,z)$. 

Casimir energy defined above, 
which has been traditionally calculated, gives $\La^5$-divergence. 
The integral $\intpE dz$ ($\int\frac{d^4p}{(2\pi)^4}dy$) appearing in 
eq.(\ref{HKA11})\ ((\ref{HK20})) corresponds to 
the summation over all positions in 5 dim bulk space $\int d^4xdz$ ($\int d^4xdy$). 
The above expression says $E_{Cas}$ is the total sum of $F(r^{-1},z)$ ($F(r^{-1},y)$) over 
the bulk space positions. We notice here the $\La^5$ divergence 
comes from the fact that 
we have overlooked some proper {\it integration measure}. 
The summation, or the {\it averaging} procedure (of F) should be properly 
defined at this stage. In the present standpoint 
we regard the {\it coordinate} system $(x^a,z)$ ($(x^a,y)$) as the {\it quantum statistical} system 
and consider 
that the coordinate $x^a$ is 
the {\it quantum mechanical} variable with the extra one $z$ ($y$) as {\it Euclidean time}. The traditional  
treatment (simple summation over the set of positions) should 
be corrected by the present quantum (geometric) approach. 
We have proposed it should be done by 
the {\it path-integral} over all hypersurfaces in the bulk space ($x^a,z$) (($x^a,y$)), as described 
in Sec.\ref{CE}. Hence 
the right expression of Casimir energy is given by (\ref{intro2}). 

%In ref.\cite{SI0801,SI0812}, 
%we claim the $\La^5$-divergence comes from this 'naive' integration measure. 
%and should be replaced by some proper measure, based on close numerical 
%calculation using some trial integration measures. 
%\newpage
%\acknowledgement{
\ack
Parts of the content of this work have been already presented at 
the international conference on "Particle Physics, Astrophysics and 
Quantum Field Theory"(08.11.27-29, Nanyang Executive Centre, Singapore)\cite{SI0903Singa}, 
%YITP Workshop on "Field Theory and String Theory" (09.7.6-10, Kyoto Univ.,Yukawa Memorial Hall), 
First Mediterranean Conference on Classical and Quantum Gravity 
(09.9.14-18, Kolymbari, Crete, Greece)\cite{SI0909}, 
%RIMS-YITP Joint Workshop on 'Duality and Scale in Quantum Science'
%(09.11.4-6, Kyoto Univ., Kyoto, Japan), 
Int. Workshop on "Strong Coupling Gauge 
Theories in LHC Era"(09.12.8-11, Nagoya Univ., Nagoya, Japan)\cite{SI0912}, 
%KEK Theory Workshop 2010 (10.3.10-13, KEK, Ibaraki, Japan), 
and 
IPMU Workshop on 'Condensed Matter Physics Meets High Energy Physics'
(10.2.8-12, IPMU, Univ. of Tokyo, Kashiwa, Japan).
%65th Japan Physical Society Meeting(10.3.20-23, Okayama Univ., Okayama, Japan). 
The author thanks the audience 
%T. Appelquist (Yale Univ.), 
%K. Fujikawa (Nihon Univ.), T. Inagaki (Hiroshima Univ.), S. Iso(KEK)
%, K. Kanaya (Univ. of Tsukuba), Y. Kitazawa(KEK), T. Kugo (Kyoto Univ.), 
%N. Sakai (Tokyo Woman's Christian Univ.), M. Sakamoto (Kobe Univ.), 
%M. Tanabashi(Nagoya Univ.), S. Watamura(Tohoku Univ.) and T. Yoneya (Univ. of Tokyo) 
for useful comments and encouragement on the occasions.

\end{document}